%% file: paper.tex
\documentclass[aps,prb,twocolumn,10pt,superscriptaddress,floatfix]{revtex4-1}

\usepackage{graphicx,epsfig,color,dcolumn}
\usepackage{ulem}
\usepackage{hyperref}
\usepackage{multirow}
\usepackage{longtable}

\begin{document}

\newcommand{\kk}{\mathbf{k}}
\newcommand{\ecoh}{E_{\scriptsize \mbox{coh}}}
\newcommand{\Exp}{\scriptsize \mbox{Exp.}}

\title{Structural and excited-state properties of oligoacene crystals from first principles}
,
\author{Tonatiuh Rangel}
\email[Email: ]{trangel@lbl.gov}
\affiliation{Molecular Foundry, Lawrence Berkeley National Laboratory, Berkeley, California 94720,USA}
\affiliation{Department of Physics, University of California, Berkeley, California 94720-7300, USA}

\author{Kristian Berland}
\affiliation{Centre for Material Science and Nanotechnology, University of Oslo,  NO-0316 Oslo, Norway}

\author{Sahar Sharifzadeh}
\affiliation{Department of Electrical and Computer Engineering and Division of Materials Science and Engineering, Boston 
University, Boston, MA 02215, USA}
\affiliation{Molecular Foundry, Lawrence Berkeley National Laboratory, Berkeley, California 94720,USA}

\author{Florian Brown-Altvater}
\affiliation{Molecular Foundry, Lawrence Berkeley National Laboratory, Berkeley, California 94720,USA}
\affiliation{Department of Chemistry, University of California, Berkeley, California 94720-7300, USA}

\author{Kyuho Lee}
\affiliation{Molecular Foundry, Lawrence Berkeley National Laboratory, Berkeley, California 94720,USA}

\author{Per Hyldgaard}
\affiliation{Department of Microtechnology and Nanoscience, MC2, Chalmers University of Technology,SE-41296 G\"oteborg, Sweden}
\affiliation{Materials Science and Applied Mathematics, Malm\"o University, Malm\"o SE-205 06, Sweden}

\author{Leeor Kronik}
\affiliation{Department of Materials and Interfaces, Weizmann Institute of Science, Rehovoth 76100, Israel}

\author{Jeffrey B. Neaton}
\affiliation{Molecular Foundry, Lawrence Berkeley National Laboratory, Berkeley, California 94720,USA}
\affiliation{Department of Physics, University of California, Berkeley, California 94720-7300, USA}
\affiliation{Kavli Energy NanoSciences Institute at Berkeley, Berkeley, California 94720-7300, USA}

\date{\today}

\begin{abstract}
Molecular crystals are a prototypical class of van der Waals (vdW)-bound organic materials with excited state properties relevant for optoelectronics applications. 
Predicting the structure and excited state properties of molecular crystals presents a challenge for electronic structure theory, as standard approximations to density functional theory~(DFT) do not capture long-range vdW dispersion interactions and do not yield excited-state properties.
In this work, we use a combination of  DFT including vdW forces-- using both non-local correlation functionals and pair-wise correction methods -- together with many-body perturbation theory~(MBPT) to study the geometry and excited states, respectively, of the entire series of oligoacene crystals, from benzene to hexacene. We find that vdW methods can predict lattice constants within 1\% of the experimental measurements, on par with the previously reported accuracy of pair-wise approximations for the same systems. 
We further find that excitation energies are sensitive to geometry, but if optimized geometries are used MBPT can yield excited state properties within a few tenths of an eV from experiment. 
We elucidate trends in MBPT-computed charged and neutral excitation energies across the acene series and discuss the role of common approximations used in MBPT.
\end{abstract}

\maketitle

\section{Introduction}

Organic solids are promising candidates for optoelectronics applications due to their strong absorption, chemical tunability, flexibility, and relatively inexpensive processing costs, among other reasons. The acene crystals, a specific class of organic semiconductors, are well-characterized, known to possess relatively high carrier mobilities,\cite{jurchescu_interface-controlled_2007,*klauk_high-mobility_2002,*stadlober_high-mobility_2005,*cheng_three-dimensional_2003} and exhibit a propensity for unique excited-state transport phenomena, notably singlet fission~(SF).\cite{hanna_solar_2006,thompson_energy_2014,congreve_external_2013,
tabachnyk_resonant_2014,bardeen_triplet_2014,chan_quantum_2013} The larger acenes in particular have received recent attention because SF was reported to be exothermic, or nearly so, for tetracene, pentacene, and hexacene.\cite{cicoira_morphology_2005,anthony_larger_2008,
smith_recent_2013,*smith_cr_2010,lee_singlet_2013,
busby_multiphonon_2014}

The interesting optoelectronic properties of acene crystals, combined with the potential for materials design via functionalization at the monomer level, have generated significant fundamental theoretical interest in these systems. Theoretical studies of excited state properties of acene crystals have often been performed with small molecular clusters, using wavefuction-based methods~\cite{zimmerman_jacs_2011,beljonne_charge-transfer_2013,
berkelbach_microscopic_2013,chan_quantum_2013,renaud_mapping_2013,
zimmerman_correlated_2013,coto_low-lying_2015},
or with extended systems, using density functional theory (DFT) and many-body perturbation theory (MBPT).\cite{tiago_ab_2003,hummer_ab_2003,*hummer_lowest_2004,
*hummer_electronic_2005,neaton_renormalization_2006,
ambrosch-draxl_role_2009,sharifzadeh_quasiparticle_2012,cudazzo_excitons_2012,*cudazzo_prb_2013,
Refaely13,*Refaely15,cudazzo_exciton_2015}
These calculations have often yielded excellent agreement with experiment and new insights into excited-state properties of acene crystals.

\begin{figure}[ht]
\includegraphics[width=0.5\textwidth]{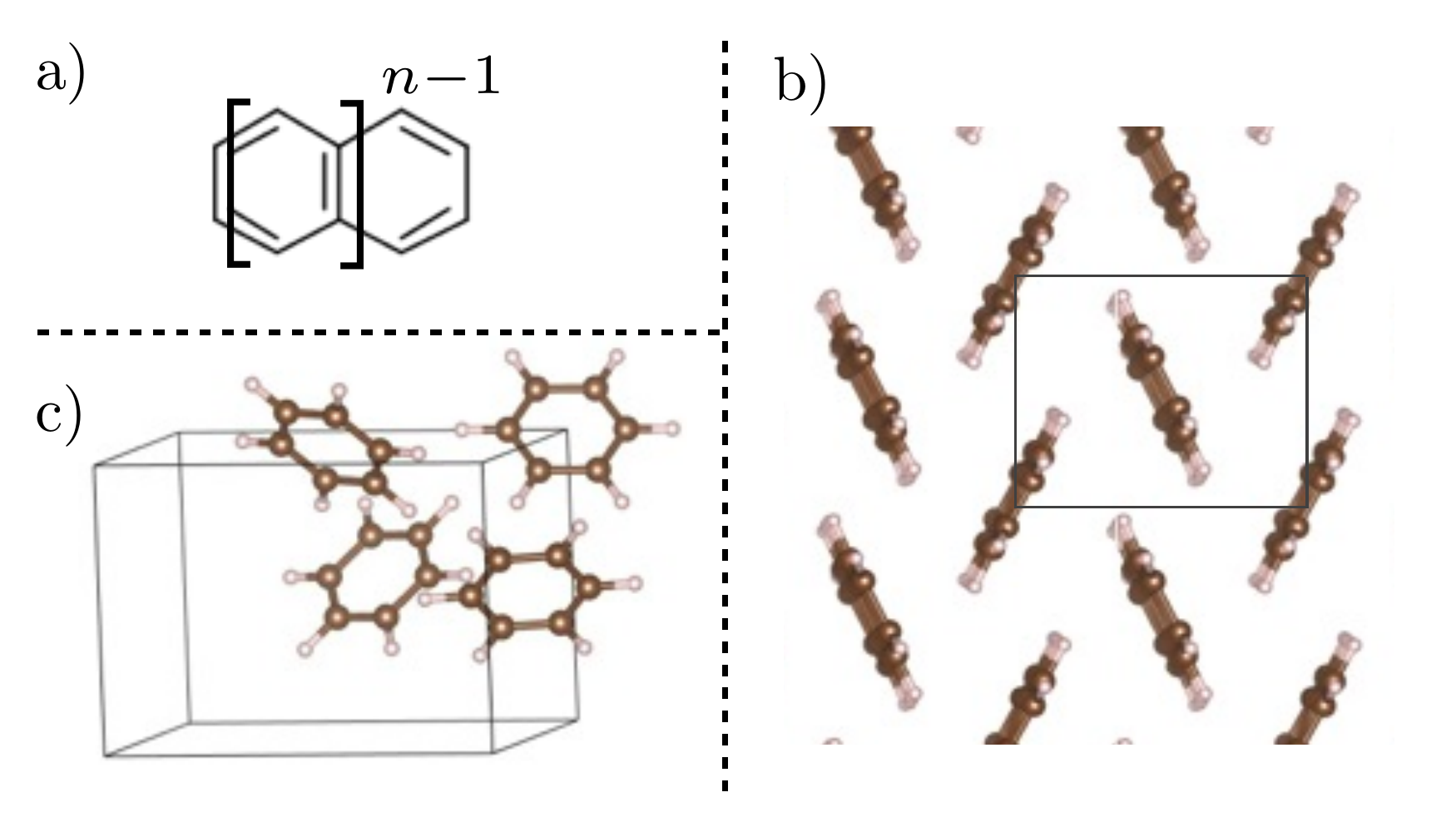}
\caption{(Color online)
The acene family. a) General formula. 
b) Herringbone structure, taken up by most acenes in the solid state, with space
group {\it P21/a} for naphthalene and anthracene and {\it P1} for larger acenes.
c) Benzene crystallizes in an orthorhombic unit cell 
with four molecules per unit cell, with space
group {\it Pbca}.}
\label{acenes}
\end{figure}

As shown in Fig.\ \ref{acenes}, acene crystals consist of aromatic monomers packed in ordered arrangements. Their constituent monomers possess strong intramolecular covalent bonds, but weak intermolecular dispersive interactions govern the crystal structure. Because the approximate exchange-correlation functionals most commonly used in DFT calculations do not account for dispersive interactions, the above-mentioned theoretical calculations have nearly always made use of experimental data for intermolecular distances and orientation.
This limits predictive power, because experimental lattice parameters can be scarce or conflicting. In particular, different polymorphs of the same material may exist, sometimes even coexisting in the same sample.
\cite{faltermeier_optical_2006,mattheus_polymorphism_2001,venuti_phonons_2004,valle_organic_2004,kakudate_polymorphism_2007,farina_pressure-induced_2003,mattheus_identification_2003,ambrosch-draxl_role_2009}
\cite{cudazzo_exciton_2015}

Fortunately, the last decade has seen rapid development of DFT-based methods that can capture dispersive interactions and several studies have demonstrated that addressing these interactions allows for predicting accurate geometries and cohesive energies of molecular solids in general and acenes in particular -- see, e.g., Refs.\ \onlinecite{nabok_cohesive_2008,ambrosch-draxl_role_2009,berland10p134705,berland_molcrys2011,otero-de-la-roza_benchmark_2012,AlSaidi12,Bucko13,kronik_understanding_2014,reilly_understanding_2013,lu_2009,jr_many-body_2014,reilly_van_2015,yanagisawa_recent_2015,Sutton15}.
Specifically, Ambrosch-Draxl {\it et al.}~\cite{ambrosch-draxl_role_2009} have suggested that a combination of dispersion-inclusive DFT methods -- which they found to predict lattice parameters in agreement with experiments for acene crystals -- followed by MBPT calculations, can be used to explore quantitative differences in optical properties of pentacene polymorphs. Their work suggests that a broader study of the entire acene family with MBPT methods, especially their recent refinements, would be highly desirable.

In this article, we combine dispersion-inclusive DFT and MBPT to study the geometry and excited states of the entire series of acene crystals, from benzene to hexacene. In each case, we compare the computed geometry, electronic structure, and optical excitations with experiment, for both the gas-phase and solid-state. To account for long-range vdW dispersive interactions, we use primarily non-local vdW density functionals~(vdW-DFs), but also employ Grimme "D2" pair-wise corrections\cite{grimme2} and compare our results where possible with previously reported data computed with the Tkatchenko-Scheffler~(TS)~\cite{ts09} pair-wise correction approach.\cite{otero-de-la-roza_benchmark_2012,schatschneider_understanding_2013}
We find that the new consistent-exchange (cx) vdW density functional~(vdW-DF-cx)~\cite{berland_exchange_2014,bearcoleluscthhy14} can predict acene lattice parameters within 1\% of low-temperature measurements, as can the TS method.
For optimized acene crystal structures, our MBPT calculations within the GW approximation and using the Bethe-Salpeter equation approach lead to gas-phase ionization potential energies, solid-state electronic band structures, and low-lying singlet and triplet excitations in good quantitative agreement with experiments. For larger acene crystals, we demonstrate that a standard G$_0$W$_0$ approach based on a semi-local DFT starting point
is insufficient, and that eigenvalue-self-consistent GW calculations are required.
Interestingly, we find that low lying excited states are sensitive to crystal geometry, particularly so for singlets, which are significantly more delocalized than triplets.
This work constitutes a comprehensive survey and validation study of both crystal structure and excited state electronic structure for this important class of molecular crystals. Furthermore, it suggests strategies for accurate predictive modeling and design of excited states in less-explored molecular systems, using current state-of-the-art methods.

The manuscript is organized as follows.
First, we summarize the computational methods used in this work in Section.~\ref{background}. Next, in Section.~\ref{sect:geom-lattice} we provide a detailed account of our calculations for the structural properties of the acene crystals, demonstrating and reviewing the accuracy of several different vdW-corrected DFT methods. We then turn to presenting MBPT results for charged and neutral excitations.
We start with charged and neutral excitations in gas-phase acene molecules, given in Section~\ref{sect:gas-phase}, followed by similar results for the solid-state in Sections~\ref{sect:electronic} and \ref{sect:opt-properties}, where we provide calculations for charged and neutral excitations, respectively, at the experimental Geometry. In Section~\ref{sect:image-charge} we critically examine the sensitivity of GW and GW-BSE calculations to structures optimized with different DFT-based approaches. Finally, we present conclusions in Section~\ref{conclusions}.

\section{Computational Methods}\label{background}
\subsection{Treatment of dispersive interactions}
\label{sect:treatvdw}

As mentioned above, great strides have been made over the past decade in the treatment of dispersive interactions within DFT -- see, e.g., Refs.\ \onlinecite{klimes_perspective:_2012,Riley12} for overviews. Of the many approaches suggested, one commonly used method is the augmentation of existing (typically semi-local or hybrid) exchange-correlation (xc) functionals by pairwise corrections to the inter-nuclear energy expression, which are damped at short range
but provide the desired long-range asymptotic behavior.\cite{Gianturco97, Johnson05,grimme1,grimme2,grimme3,ts09,Otero13,Corminboeuf14} The most widely used examples of this idea are the D2\cite{grimme2} and D3\cite{grimme3} corrections due to Grimme and the Tkatchenko-Scheffler (TS)~\cite{ts09} correction scheme. A different commonly used approach, known as vdW-DF, includes dispersion interactions via an explicit non-local correlation functional.\cite{Langreth09,Hyldgaard:Interpret2014,berland_van_2015}
Several vdW-DF versions are in use, starting with the original vdW-DF1 \cite{Dion} functional. These include, e.g., an improved version, vdW-DF2\cite{lee10p081101}, making use of a more accurate semilocal exchange functional and an updated vdW kernel; the simplified yet accurate form of Vydrov and van Voorhis, VV10 \cite{VV10}; and the more recently developed vdW-DF-cx \cite{berland_exchange_2014} functional, an update with improved performance for lattice constants and bulk moduli of layered materials and dense solids. In the following, we abbreviate vdW-DF1 as DF1, etc., for functionals in the vdW-DF class.

\subsection{Many-body perturbation theory}
\label{sect:MBPT}

As mentioned above, our first principles MBPT calculations are based on the GW approach for charged excitations and on the GW-BSE approach for neutral ones. GW calculations proceed pertubatively based on a DFT starting point, which for solids is usually computed using the Kohn-Sham equation within the local density approximation (LDA) or the generalized gradient approximation (GGA).   
The Kohn-Sham eigenvalues and eigenfunctions are used to evaluate approximatel, the self-energy operator, $\Sigma$, as $iGW$, where $G$ is the one-electron Green function of the system and $W=\epsilon^{-1} v$ is the dynamically screened Coulomb interaction; $v$ is the Coulomb potential and $\epsilon$ is the wave-vector and frequency-dependent dielectric function.\cite{hybertsen_electron_1986,hedin_new_1965} 
The DFT eigenvalues are then updated via first-order perturbation theory. This approach is known as the G$_0$W$_0$ approximation. 
This method is often very successful, but nevertheless it is somewhat dependent on the DFT starting point. 
GW can be evaluated, in principle, self-consistently by different approaches~\cite{marom_benchmark_2012,rostgaard_fully_2010,korbel_benchmark_2014,shishkin_self-consistent_2007,vanschilfgaarde_quasiparticle_2006, marom_benchmark_2012,bruneval_effect_2006,
schindlmayr_spectra_1998}, mitigating the starting point dependence by iterating over eigenenergies and eigenvalues. 
Given the computational demands associated with acene crystals, in the following we limit our study to the diagonal part of $\Sigma$ and, if going beyond G$_0$W$_0$, we only update the eigenvalues in $G$ and $W$, retaining the original DFT wavefunctions under the assumption that they are close to the true QP wavefunctions.\cite{hybertsen_electron_1986,luo_quasiparticle_2002,Blase11b,faber_excited_2014} 
We denote this sort of partial self-consistency as evGW, where ``ev'' emphasizes that self-consistency is achieved only with respect to the eigenvalues.

Given the GW-computed quasi-particle energies, as well as the static inverse dielectric function computed within the random phase  approximation, we compute neutral excitation energies 
by solving the Bethe-Salpeter equation (BSE).\cite{Onida95,Onida02,Rohlfing00} We use an approximate form of the BSE, developed within a first principles framework by Rohlfing and
Louie,\cite{Rohlfing00} which involves solving a new eigenvalue problem obtained from an electron-hole interaction matrix. We perform the solution within the Tamm-Dancoff approximation (TDA) and limit our calculations to low-lying singlet and triplet excitations. 

\subsection{Computational details}
\label{sect:tec-dets}

Our DFT calculations are performed with the \textsc{Quantum Espresso}~(QE) package~\cite{qe}, unless otherwise indicated.  
$\Gamma$-centered Monkhorst-Pack k-point grids are used for all calculations.~\cite{monkhorst_special_1976}
For geometry optimizations, where Hellmann-Feynman forces and stress tensor components are minimized, we use a number of $\kk$-points along each crystallographic direction corresponding to a spacing of $\sim$3.3~Bohr$^{-1}$ between neighboring points in reciprocal space. 
All Hellmann-Feynman forces are converged to 10$^{-5}$~Ry/Bohr and total energies are converged to 10$^{-5}$~Ry.
We use a plane-wave basis kinetic energy cutoff of 55 Ry. Taken together, these choices lead to total energies converged to 1~meV per atom.

For calculations with vdW-DF functionals, we use the ultrasoft pseudopotentials (USPPs) given in Ref.~\onlinecite{berland_exchange_2014};
for vdW approaches based on inter-atomic pairwise potentials, we use Fritz-Haber-Institut~(FHI) norm-conserving~(NC) pseudopotentials~(PPs),~\cite{fuchs_ab_1999} because these corrections are not compatible with USPPs in the present version of QE.
Following a prior successful approach with vdW density functionals,\cite{callsen_assessing_2015}  we use Perdew-Burke-Ernzerhof (PBE)~\cite{Perdew96} PPs for DF2 and DF and PBEsol~\cite{Perdew08} PPs for DF-cx.\cite{berland_van_2015}
In principle, native vdW-PPs have begun to be explored with vdW-DFs, and we relegate the evaluation of such pseudopotentials for acenes to future work.\cite{hamada_pseudopotential_2011}
The latter choice is based on the fact that the exchange functional of DF-cx is much closer in form to PBEsol than to PBE.
A test study reveals that the results are not significantly affected by this choice: for naphthalene, the lattice parameters (and volume) obtained using DF-cx with PBE PPs differ by no more than 1.2\% (0.2\%) from standard DF-cx calculations.

To test the reliability of our PP choice, we benchmarked our calculations of solid naphthalene (see Section~\ref{sect:geom-lattice} below for details) against other codes and pseudopotentials. 
The lattice parameters obtained with our USPPs, the FHI NC-PPs available at the QE site~\cite{qe-site}, and 
Garrity-Bennett-Rabe-Vanderbilt (GBRV)~\cite{garrity_pseudopotentials_2014} USPPs agree within 0.3\%. Additionally, we relaxed the structure of benzene with the \textsc{VASP} code, using projector-augmented waves ~\cite{kresse_ultrasoft_1999} with vdW-DF2, obtaining lattice parameters in agreement with those obtained from Quantum Espresso to within 0.4\%.
Note that a higher, 110 Ry cutoff was used for the FHI-NC-PPs calculations. 
The GBRV-USPPs were constructed to be exceptionally hard and required a plane wave cutoff of 350 Ry to achieve a convergence threshold of 1 meV/atom. 

For each acene crystal, using any of the DFT approximations mentioned above, following geometry optimization we compute cohesive energies~($\ecoh$) via the standard relation,
\begin{equation}
\ecoh=E^{\rm gas}-\frac{1}{N} E^{\,\rm solid},
\end{equation}
where $E^{\rm gas}$ is the total energy of an isolated monomer, $E^{\rm solid}$ is the total energy of the solid phase unit cell, and $N$ is the number of molecules per unit cell in the solid.

Our MBPT calculations are performed with the \textsc{BerkeleyGW} package.~\cite{berkeleygw} 
Capitalizing on its efficient and highly-parallel diagonalization techniques, Kohn-Sham starting-point wavefunctions and eigenenergies for input into MBPT are generated with the \textsc{ABINIT} software suite.~\cite{abinit}

In some of the calculations given below, we deliberately use experimental lattice constants to study the accuracy of the GW-BSE appproach independent of geometry. For consistency, we use room-temperature experimental data for all acenes\cite{bacon_crystallographic_1964,schiefer_determination_2007,
capelli_molecular_2006,mason_crystallography_1964,campbell_crystal_1962}
except for hexacene, where crystallographic data are only available at $T=123~K$~\cite{watanabe_synthesis_2012}. For pentacene, we simulate the thin-film polymorph (denoted below as P$_3$), because it is the one most commonly measured in experiment (see Sect.\ref{sect:geom-lattice}). In other calculations, meant to explore the impact of the geometry, we use the optimized geometry obtained from the DFT calculation.

We note that \textsc{BerkeleyGW} requires NC-PPs as input, but we use USPPs for lattice optimizations. Prior to the MBPT calculations, we relaxed the internal coordinates using NC-PPs within PBE, with the lattice parameters held fixed at their optimized value. This was found to result in negligible differences for both geometry and excited state properties.
We followed the same internal relaxation procedure when using experimental lattice vectors, following Ref.~\onlinecite{sharifzadeh_quasiparticle_2012}.

Our GW calculations involve a number of convergence parameters, which are set to assure that quasiparticle gaps, highest-occupied molecular orbitals~(HOMOs), and band edge energies for crystals and gas-phase molecules are converged to $\sim$0.1~eV. 
Our dielectric function is extended to finite frequency using the generalized plasmon-pole~(GPP) model of Hybertsen and Louie,~\cite{hybertsen_electron_1986}, modified to handle non-centrosymmetric systems by Zhang {\it et al}.~\cite{zhang_evaluation_1989}
For solids, we use an energy cutoff of 10~Ry to truncate the sums in {\bf G}-space used for the calculation of the polarizability. 
We sum over a number of unoccupied bands equivalent to an energy range of 30~eV. 
Response functions and $\Sigma$ are evaluated on $\kk$-point meshes selected to lead to a spacing of $\sim$1.6 Bohr$^{-1}$ in reciprocal space. 
For gas-phase molecules, we use an energy cutoff of 25~Ry for the polarizability  and sum over a number of unoccupied bands equivalent to 52~eV above the lowest unoccupied molecular orbital~(LUMO) energy. 
Molecules are modeled in a large supercell with dimensions chosen to contain 99\% of the HOMO (see Supplemental Material for details), with the internal coordinates relaxed using PBE.
We use the static-remainder technique to accelerate the convergence with number of bands,~\cite{tiago_optical_2006} using the version of Deslippe {\it et al}.~\cite{deslippe_coulomb-hole_2013} 
A Wigner-Seitz Coulomb truncation scheme is used to eliminate interactions between molecules of neighboring cells in the periodic lattice.\cite{berkeleygw}
These convergence criteria and parameters have been tested and used in Ref.~\onlinecite{sharifzadeh_quantitative_2012}.

For our BSE calculations, the BSE coupling matrix is constructed with 8~valence $\times$ 8~conduction bands, sufficient to converge the transition energies involving the lowest states, as shown explicitly in the supplemental material.
Two $\kk$-point meshes are used: a coarse $\kk$-point mesh for the BSE kernel and a fine $\kk$-point mesh to calculate the low-lying excited states.  
Coarse k-meshes are chosen to be the same as those used in the GW step, while fine meshes are the same as in the geometry optimization. 
These k-meshes are explicitly provided in the Supplemental Material.

\section{Results and discussion}

\begin{figure*}[t]
\begin{tabular}{@{}cc@{}} 
  \includegraphics[width=0.45\linewidth]{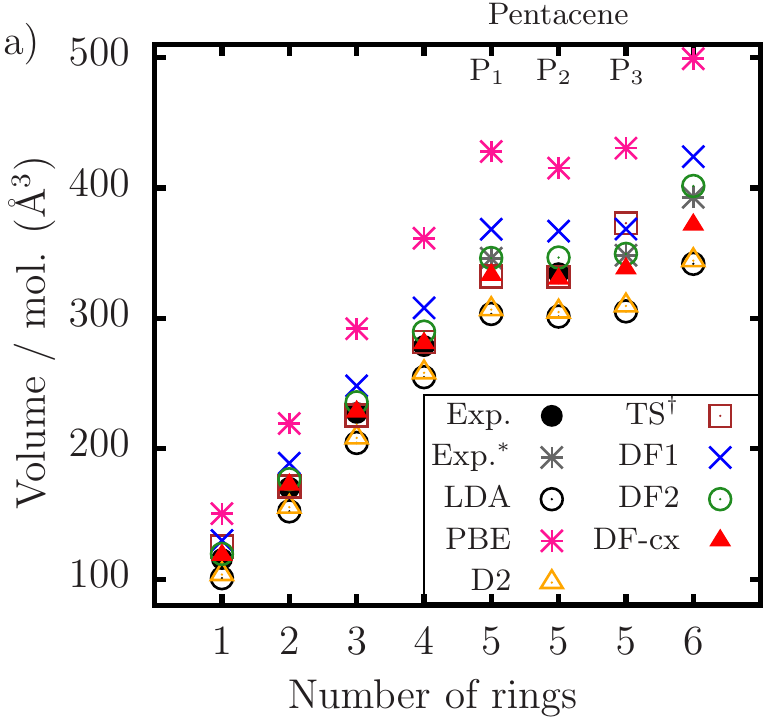}
\hspace{0.15in}
 \includegraphics[width=0.45\linewidth]{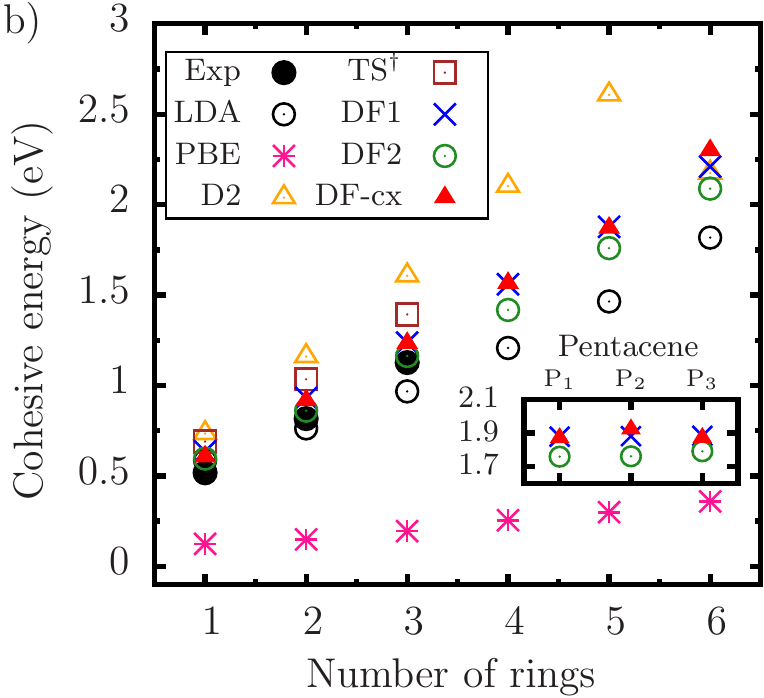}
\end{tabular}
\caption{(Color online)
(a) Volume per molecule for the acene crystals, calculated using different approximations within DFT  --
LDA (black empty-circles),
PBE (pink stars),
DF1 (blue crosses),
DF2 (green empty-circles), 
DF-cx (red filled-triangles),
PBE-D2 (orange empty-triangles), and PBE-TS (brown squares).
These are compared to low temperature experimental data, for T$\leq 16$~K from Refs.~\onlinecite{david_crystal_1992,chaplot_structure_1982,capelli_molecular_2006} and extrapolated to $0$~K as indicated in Appendix~\ref{appendix:vdws} (in black filled-circles).
For two pentacene polymorphs and hexacene, only experimental data at T$\geq90$~K is available~\cite{campbell_crystal_1962,watanabe_synthesis_2012,schiefer_determination_2007} (in dark-grey stars).
(b) Cohesive energies~$\ecoh$ for the acene series, obtained with the same set of approximations as in (a). 
Experimental $\ecoh$ (black filled-circles) are obtained from enthalpies of sublimation (Ref.~\onlinecite{chickos_enthalpies_2002} see text).
Inset: calculated $\ecoh$ for three pentacene polymorphs.
$\dagger$ PBE-TS cohesive energies are taken from Ref.~\onlinecite{reilly_understanding_2013} and PBE-TS volumes from Refs.~\onlinecite{otero-de-la-roza_benchmark_2012,schatschneider_understanding_2013}.
}
\label{fig:vdws}
\end{figure*}

\subsection{Lattice Geometry and Cohesive Energy}
\label{sect:geom-lattice}

We begin our discussion by considering the effect of the chosen DFT approximation on the crystal geometry and cohesive energy. Experimental unit cell volumes for the acene crystals are compared in Fig.~\ref{fig:vdws}a with volumes calculated using the LDA, PBE, PBE-D2, PBE-TS (from Refs.\ \onlinecite{otero-de-la-roza_benchmark_2012,schatschneider_understanding_2013}), DF1, DF2, and DF-cx approaches. A similar comparison for cohesive energies is given in Fig.~\ref{fig:vdws}b. A complete set of structural data, along with error estimates, is given in Appendix~\ref{appendix:vdws}. 
For tetracene, its polymorph 1 (P$_1$) also called high-temperature polymorph,\cite{campbell_crystal_1962,holmes_nature_1999} referred to as \textsc{TETCEN} in the \textsc{Cambridge Structural Database}~(CSD),\cite{csd} is considered. 
This crystal is known to undergo a pressure-assisted transition to a different high-pressure or low-temperature polymorph (P$_2$),\cite{venuti_phonons_2004,kalinowski_pressure_1978,kolendritskii_exciton_1979,rang_hydrostatic-pressure_2001,sondermann_x-ray_1985,oehzelt_crystal_2006}
the study of which is beyond the scope of this work. This low-temperature polymorph has been successfully described within the TS method in Ref.\ \onlinecite{schatschneider_understanding_2013}.
For pentacene, three well-known polymorphs are considered, using experimental structures available in the CSD.\cite{csd} These are:
\begin{itemize}
\item
P$_1$: the Campbell structure, referred to as \textsc{PENCEN} in the CSD. It is also known as the high-temperature polymorph.
Found first by Campbell in 1962,~\cite{campbell_crystal_1962} it had been lost until reported again in 2007.~\cite{siegrist_polymorph_2007}
\item P$_2$: a common bulk-phase polymorph, referred to as \textsc{PENCEN04} 
in the CSD.\cite{mattheus_polymorphism_2001,yoshida_crystallographic_2008}
\item P$_3$: a common thin-film polymorph, referred to as \textsc{PENCEN10} in the CSD.~\cite{schiefer_determination_2007,yoshida_crystallographic_2008} Most experimental data correspond to this polymorph.
\end{itemize}

Fig.\ \ref{fig:vdws} shows, as expected, that standard (semi-)local functionals do not agree well with experimental results. PBE significantly overestimates lattice constants and underestimates cohesive energies. This can be attributed directly to the lack of treatment of dispersive interactions in PBE. \cite{kronik_understanding_2014} LDA lattice constants are underestimated by $\sim$3\%, but this binding is spurious, rather than reflecting a successful treatment of dispersive interactions.\cite{kronik_understanding_2014} The spurious binding is attributable to the insufficient treatment of exchange.~\cite{murray_investigation_2009,berland_analysis_2013}

Turning to explicit vdW functionals, Fig.\ \ref{fig:vdws}a clearly shows that DF1 overestimates lattice constants essentially as much as LDA underestimates them. This is because DF1 is based on the exchange of revPBE,\cite{revPBE} a variant of PBE with exchange that is too repulsive for the systems studied here. At the same time, Fig.\ \ref{fig:vdws}b shows that it still overestimates binding energies. 
We note that cohesive energies of acene crystals have been calculated with DF1 prior to this work,\cite{otero-de-la-roza_benchmark_2012,ambrosch-draxl_role_2009,nabok_cohesive_2008} with differing conclusions. While DF1 results for $\ecoh$ are in agreement with experiment to better than 5\% in Refs.~\onlinecite{ambrosch-draxl_role_2009,nabok_cohesive_2008}, Ref.~\onlinecite{otero-de-la-roza_benchmark_2012} reports DF1 results that deviate from experiment by as much as $\sim 17\%$. These differences can be partially explained by the different choices these studies made for the experimental reference data. Some differences remain even if we use the 
experimental values of Ref.~\onlinecite{reilly_understanding_2013}, in which the contributions due to vibrations are carefully taken into account, throughout. Despite having carefully ruled out lack of convergence in our calculations, the average percentage error~(see Table.~\ref{table:cohesive} in the Appendix) in $\ecoh$ is then somewhat larger in the present study, being 16\%, 10\% and 9\% in the data of our work, Ref.~\onlinecite{otero-de-la-roza_benchmark_2012}, and Ref.~\onlinecite{ambrosch-draxl_role_2009}, respectively. For the lattice parameters, however, we find good agreement (within $2$\%) with those reported previously.

Fig.\ \ref{fig:vdws}a clearly shows that DF2 improves geometries with respect to DF1, in agreement with the findings in Ref.~\onlinecite{otero-de-la-roza_benchmark_2012}, with further improvement gained from DF-cx. Specifically, lattice constants are within 2\% and 1\%, respectively, of experiment.
Fortuitously, DF2 values for the lattice parameters are similar to the thermally expanded lattice parameters obtained at room temperature. 
This is attributable to a cancellation of errors, as we model the structure at zero Kelvin.
Recent work~\cite{yanagisawa_recent_2015} reported that a DF2 variant, called rev-vdW-DF2,\cite{hamada_van_2014} predicts lattice constants 
for benzene, naphthalene, and anthracene that are in remarkable agreement with low temperature experiments (within 0.5\%). 
For tetracene and P$_2$ pentacene, good agreement with room temperature experiments is found,\cite{yanagisawa_recent_2015} 
but the reported volumes overestimate structures extrapolated to zero Kelvin by $\sim$2\% for pentacene P$_2$ and 8\% for tetracene.

For cohesive energies, Fig.\ \ref{fig:vdws}b shows that neither DF2 nor DF-cx improve meaningfully upon DF1 cohesive energies. Specifically, the values obtained for DF2 are in excellent agreement (within 0.05~eV) with those reported in Ref.~\onlinecite{otero-de-la-roza_benchmark_2012}, as is the conclusion regarding lack of improvement over DF1. Interestingly, rev-vdW-DF2 reduces the error in cohesive energies with respect to experiments by half. \cite{yanagisawa_recent_2015} 

Turning to pair-wise correction methods, Fig.\ \ref{fig:vdws}a shows that lattice vectors calculated with D2 and TS corrections, added to an underlying PBE calculations, are within 3\% and 1\% of experimental data, respectively, whereas cohesive energies are within 30\% to 40\% of experiment. Thus, they perform as well as DF methods in terms for geometries prediction but somewhat worse for cohesive energies.

To summarize, both the latest pair-wise approaches and the latest DF methods can provide lattice parameters in outstanding agreement with experimental data (within $\sim$1\%) across the acene series, illustrating the predictive power of vdW methods and allowing for an excellent geometrical starting point for MBPT calculations. However, errors in cohesive energy are still on the order of 10\% to 30\%.
In future work, it would be interesting to examine whether techniques which add non-locality beyond pair-wise interactions, particularly the many-body dispersion method \cite{jr_many-body_2014,reilly_understanding_2013} can reduce the error in the cohesive energy. It would also be interesting to examine Grimme's ``D3'' method \cite{grimme3}, which also attempts to mimic many-body terms and other features that may improve calculated lattice constants and energies with respect to the ``D2'' approach.~\cite{ehrlich_dispersion-corrected_2012}

\subsection{Charged and neutral excitations of gas-phase molecules}
\label{sect:gas-phase}

\begin{figure}[h!t]
\includegraphics{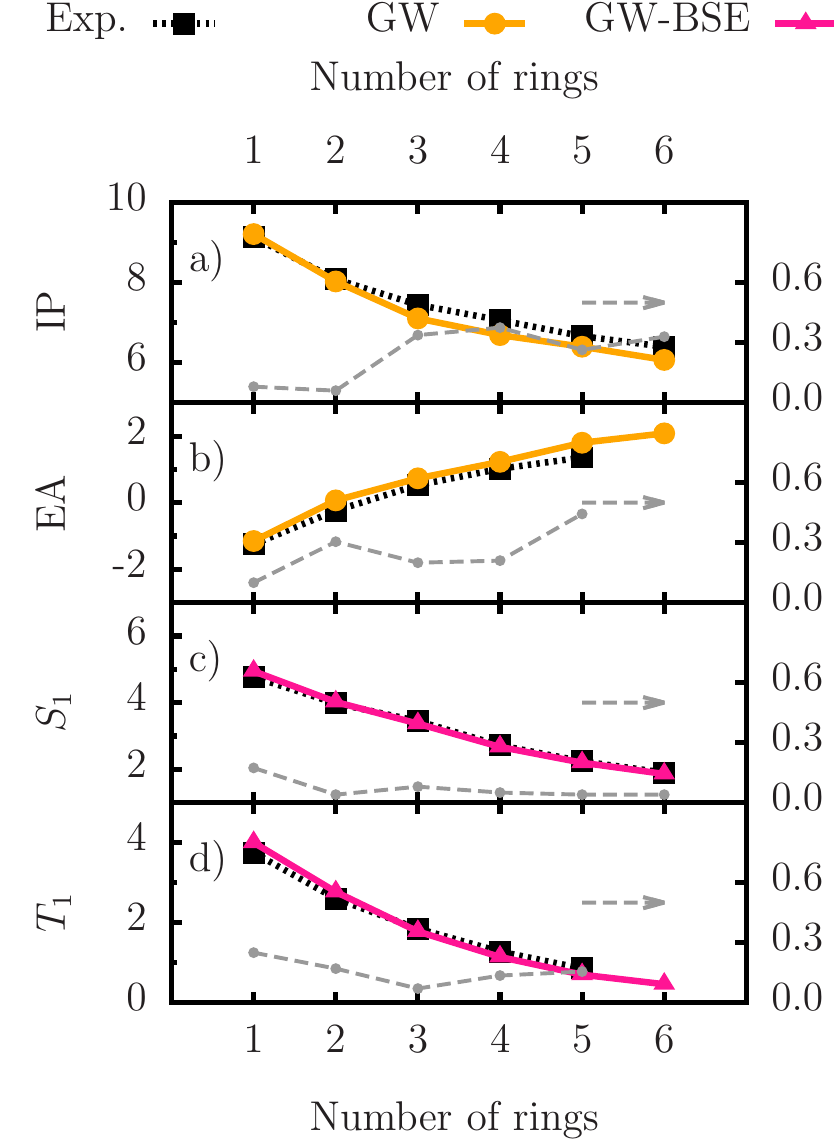}
\caption{(Color online) 
Excited-state energetics, in eV, of the gas-phase acene molecules.
(a) Ionization potentials~(IPs) and 
(b) electron affinities~(EAs) calculated within  
GW (solid orange [light-grey] lines [circles]),
as well as the (c) lowest singlet~($S_1$)  
and (d) lowest triplet~($T_1$) excitation energies 
calculated within 
GW-BSE (solid pink [grey] lines [triangles]).
All results are compared with experimental
data (dashed black lines [squares]),~\cite{frank_unoccupied_1988,pope_electronic_1999,
*clark_-orbital_1972,*malloci_time-dependent_2007,
*halasinski_electronic_2000,*salama_electronic_1991,
doering_electronic_1977}.
 The absolute deviation from experiment is given in eV in grey dashed-lines~(right axis). As discussed in the text, G$_0$W$_0$ based on a PBE starting point and the GPP approximation is used throughout.}
\label{fig:energetics-gas-phase}
\end{figure}

\begin{table}
\label{table:gas_phase}
\setlength\extrarowheight{2pt}
\begin{ruledtabular}
\begin{tabular}{llcccccc}
&&\multicolumn{6}{c}{Number of rings}\\
&& 1 & 2 & 3 & 4 & 5 & 6\\
 \hline \multirow{2}{*}{IP}
  & GW &
9.2 & 8.0 & 7.1 & 6.6 & 6.4 & 6.1\\
   & Exp. &
9.0 -- 9.3 & 8.0 -- 8.2 & 7.4 & 7.0 -- 7.2 & 6.6 & 6.4 \\
\hline \multirow{2}{*}{EA}
  & GW &
-1.2 & 0.1 & 0.7 & 1.2 & 1.8 &2.1\\
   & Exp. &
-1.2 -- -1.4 & -0.2 & 0.5 & 1.0 & 1.3 & \\
\hline \multirow{2}{*}{$S_1$}
 &GW-BSE &
4.9 & 4.0 & 3.4 & 2.7 & 2.2 & 1.9\\
 &Exp.&
4.8 & 4.0 & 3.5 & 2.7 & 2.2 & 1.9\\
\hline \multirow{2}{*}{$T_1$}
 &GW-BSE &
4.0 & 2.8 & 1.8 & 1.1 & 0.7 & 0.5\\
 &Exp.&
3.7 & 2.6 & 1.9 & 1.3 & 0.9 &
\end{tabular}
\end{ruledtabular}

\caption{Charged and neutral excitation energies for gas-phase acene molecules. 
Theoretical and experimental ionization potential~(IP), electron affinity~(EA), lowest singlet~($S_1$) and lowest triplet~($T_1$) energies are tabulated, in eV. IP and EA are calculated within G$_0$W$_0$, as described in the text. $S_1$ and $T_1$ are calculated within the G$_0$W$_0$-BSE approach.
Experimental data are taken from Refs.~\onlinecite{frank_unoccupied_1988,pope_electronic_1999,*clark_-orbital_1972,*malloci_time-dependent_2007,*halasinski_electronic_2000,*salama_electronic_1991,doering_electronic_1977}.
}
\label{table:energetics-gas-phase}
\end{table}

Before discussing excitations in acene solids, it is instructive to consider charged and neutral excitations in the constituent gas-phase molecules. Computed results for the ionization potential~(IP) and electron affinity~(EA), computed with the GW approach, as well as lowest-energy singlet~($S_1$) and triplet~($T_1$) excitation energies, computed within the GW-BSE approach, of gas-phase acene molecules, are given in Fig.~\ref{fig:energetics-gas-phase}. The same data are presented in Table~\ref{table:energetics-gas-phase}.

We find that calculated G$_0$W$_0$-computed IPs and EAs are within 0.4~eV of experiment, with an average error of only 0.2~eV. The agreement is particularly good for the smallest acenes,\cite{sharifzadeh_quantitative_2012} for reasons that have to do with our use of the PPM, as elaborated in Ref.\ \onlinecite{lischner_effects_2014}. 
For the largest acenes, the deviations of the IP and EA values from experiment possess opposite signs, leading to a larger error (up to 0.7~eV) in the fundamental gap, i.e., the difference between the IP and the EA. 
Many recent studies -- see, e.g., Refs.\ \onlinecite{bruneval_benchmarking_2012,marom_benchmark_2012,
KM12,faber_excited_2014,bruneval_systematic_2015,govoni_large_2015,hirose_all-electron_2015,jacquemin_benchmarking_2015} -- indicate that a different starting point for the 
G$_0$W$_0$ calculation, or use of self-consistent GW scheme, will improve agreement with experiment. Based on the results of, e.g., Refs.\ \onlinecite{sharifzadeh_quasiparticle_2012,marom_benchmark_2012,Blase11} for some of the acenes, we expect the same here, but do not pursue this point further as we wish to facilitate the comparison to the solid-state data given below.

The neutral singlet and triplet excitation energies, $S_1$ and $T_1$, computed with G$_0$W$_0$-BSE, are close to experimental values, deviating by 0.3 eV at most across the entire series. Given that, as mentioned above, the fundamental gap exhibits larger discrepancies between theory and experiment, the accuracy of the neutral excitation energies is likely to be partly due to a cancellation of errors between the G$_0$W$_0$ gaps and BSE binding energies.

\subsection{Charged excitations in acene crystals}
\label{sect:electronic}

We begin our MBPT analysis of the acene series by intentionally 
using the experimental geometries ~\cite{bacon_crystallographic_1964,schiefer_determination_2007,
capelli_molecular_2006,mason_crystallography_1964,campbell_crystal_1962,
watanabe_synthesis_2012} as our starting point. This is done to isolate errors associated with the particular flavor of the GW-BSE method used here from errors related to structural deviations (the latter are analyzed below, in Section~\ref{sect:image-charge}).

\begin{table}
\input{Figures/table6.tex}
\caption{
Fundamental gaps of the acene crystal series, computed within the G$_0$W$_0$ and evGW approximations, compared to experimental data, taken from Refs.\ \onlinecite{geacintov_lowlying_1969,braun_intrinsic_1970,baessler_hot_1972, baessler_bandgap-determination_1973, belkind_energy_1974,riga_esca_1977, sebastian_charge_1981,silinsh_charge_1982, sebastian_charge-transfer_1983,
isono_two-color_1986}.
Experimental lattice constants have been used in the calculations throughout. All quantities are in eV.
}
\label{table:g0w0-evgw-charged}
\end{table}

GW results for the fundamental gap, compared wherever possible to experiment, are summarized in Table~\ref{table:g0w0-evgw-charged}. GW bandstructures are provided in the Supplemental Material.
Table~\ref{table:g0w0-evgw-charged} shows that the 
G$_0$W$_0$ results fully capture the quantum-size effect, i.e., the reduction of the fundamental gap value with increasing acene size. Furthermore, for most acenes G$_0$W$_0$ yields fundamental gaps in good agreement with experimental data. The computed data somewhat underestimate experimental values for $n$=3 to $n$=5, an effect partly compensated for by gap reduction owing to thermal expansion in the experimental data, which was taken at higher temperatures, mostly room temperature. However, our G$_0$W$_0$ values decrease too rapidly with size. Thus for naphthalene the fundamental gap is somewhat overestimated but for pentacene it is somewhat underestimated. For hexacene, the G$_0$W$_0$ is no longer acceptable. While we are not aware of an experimental fundamental gap value, the G$_0$W$_0$ value we compute is smaller than the singlet excitation energy (see Section~\ref{sect:opt-properties} below) and therefore certainly underestimates the fundamental gap.

As in the gas-phase data, we attribute the discrepancy in hexacene to a starting point effect. We note that for pentacene, it was shown in Ref.~\onlinecite{sharifzadeh_quasiparticle_2012} that the QP gaps obtained with the plasmon-pole model and with a full-frequency integration are essentially identical. Therefore, we do not believe that use of the plasmon-pole approximation plays a major role here. The evGW method partly compensates for starting point effects. However, as also shown in Table~\ref{table:g0w0-evgw-charged}, evGW tends to overestimate the experimental gaps. For hexacene, however, we prefer the evGW value as it offers a compensation for the underestimate of the G$_0$W$_0$-computed  value (an issue confirmed by optical data presented in Section \ref{sect:opt-properties} below). 

\begin{figure*}
\begin{tabular}{cc}    \includegraphics[width=0.5\linewidth]{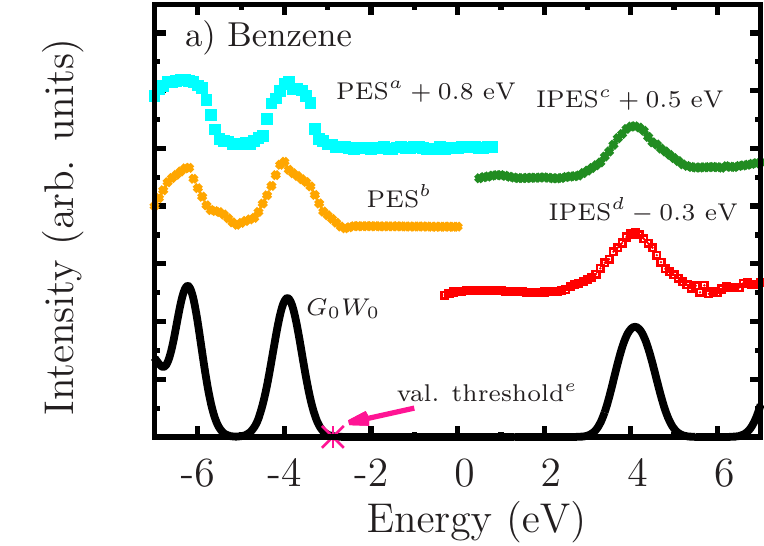}
&
\includegraphics[width=0.5\linewidth]{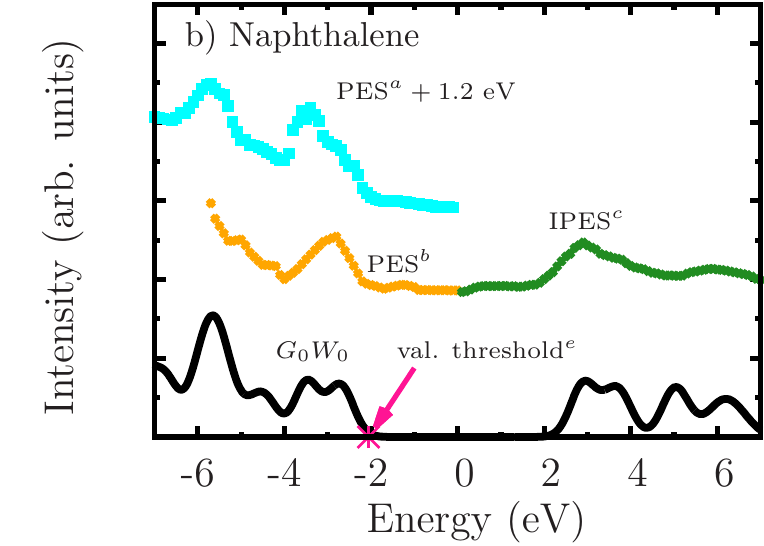}\\
\includegraphics[width=0.5\linewidth]{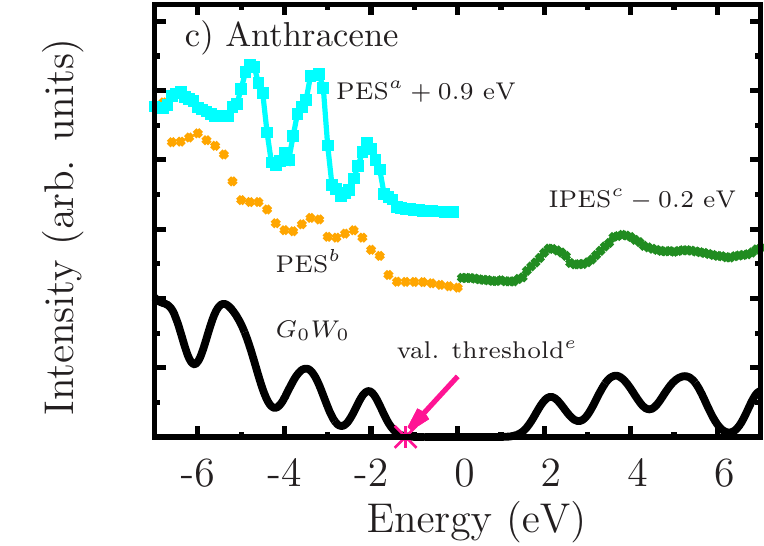}&
\includegraphics[width=0.5\linewidth]{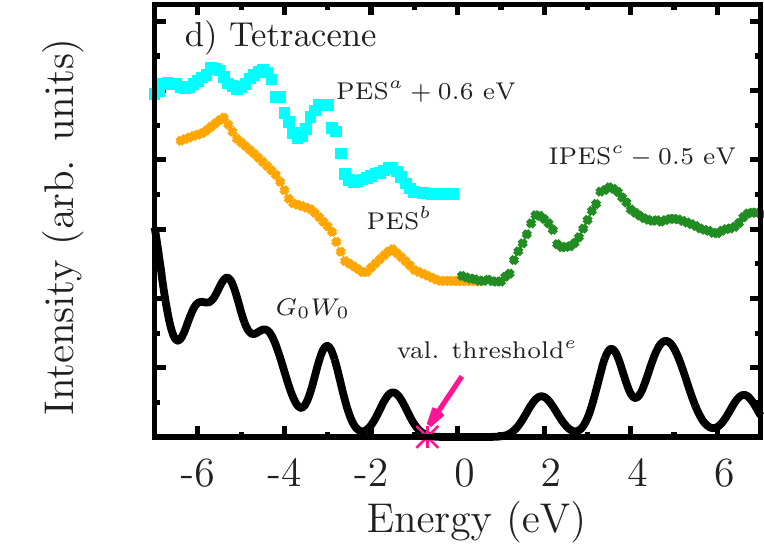}\\
\includegraphics[width=0.5\linewidth]{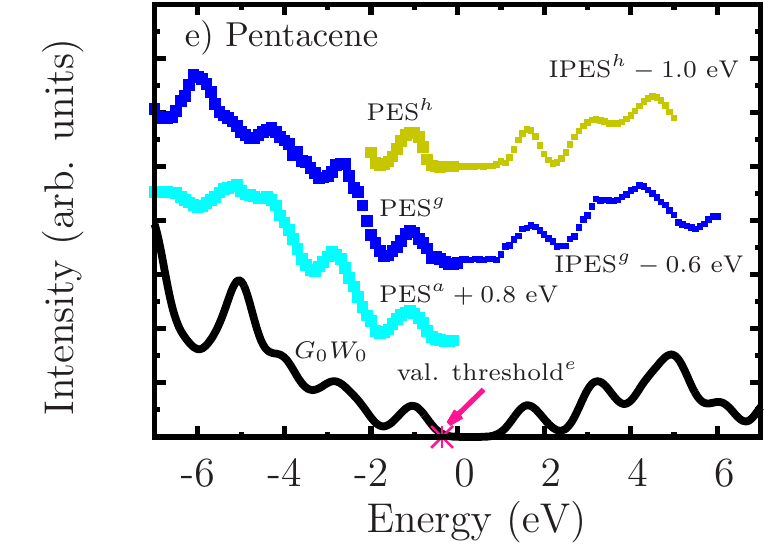}&
\includegraphics[width=0.5\linewidth]{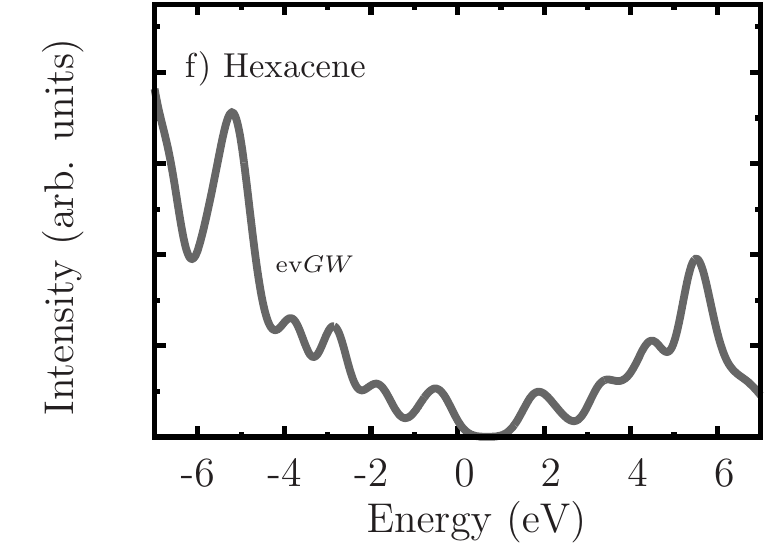}
\end{tabular}
\caption{Quasiparticle DOS, calculated using GW, compared with  experimental photoemission~(PES) and inverse photoemission~(IPES) spectra.
Two different GW approximations are used: G$_0$W$_0$ for benzene through pentacene and evGW for hexacene -- see text for details.
The calculations are based on experimental lattice parameters\cite{bacon_crystallographic_1964,schiefer_determination_2007,
capelli_molecular_2006,mason_crystallography_1964,
campbell_crystal_1962,watanabe_synthesis_2012} to avoid errors related to geometry.}
In each case, the DOS is interpolated on a dense mesh of $\kk$-points using maximally-localized Wannier functions,~\cite{marzari_maximally_1997} and broadened by convolution with a 0.4 eV Gaussian. Bandstructures and DOS with lower broadening are given in the supplemental material.
Experimental PES data have been rigidly shifted, by an amount indicated in the figure, so as to match reference ionization potential data of Refs.\ \onlinecite{asaf_photoemission_1975,williams_photoemission_1967}~(pink stars) and PES data of Ref.~\onlinecite{riga_comparative_1977}~(orange points). The GW valence band edge has been set to the same position. IPES data have then been shifted to match the GW-computed position of the conduction band edge. See text for full details.
$^a$ Ref.~\onlinecite{rocco_electronic_2008}.
$^b$ Ref.~\onlinecite{riga_comparative_1977}. 
$^c$ Ref.~\onlinecite{frank_unoccupied_1988}. 
$^d$ Ref.~\onlinecite{frank_electron_1986}. 
$^e$ Refs.~\onlinecite{asaf_photoemission_1975,williams_photoemission_1967}.
\label{pes-ipes-gw}
\end{figure*}

Beyond bandgap values, it is very instructive to compare the GW-calculated electronic density of states (DOS) to measured photoemission and inverse photoemission spectroscopy~(PES and IPES, respectively) data.
Such a comparison is not straightforward. Experimentally, it is challenging to pinpoint absolute conduction and valence band energies.\cite{Hill00,Krause08} As discussed in detail in Ref.\ \onlinecite{sharifzadeh_quasiparticle_2012}, agreement between theory and experiment is often observed only after a rigid shift (of valence and conduction bands separately). This rigid shift has been attributed to a combination of several physical effects, including surface polarization, vibrational contributions, and a dynamical lattice, and to some extent also to residual errors of both theory and experiment. We therefore employ the same rigid shift procedure here, as follows. 
First, because absolute potentials are never defined in periodic boundary calculations, we align the top of the GW-computed valence band with experimental values from Refs.~\cite{asaf_photoemission_1975,williams_photoemission_1967}. To compare with experiment, each photoemission and inverse photoemission curve is aligned with the GW valence and conduction band DOS, respectively. 
Based on the results of Table~\ref{table:g0w0-evgw-charged}, this procedure is performed using G$_0$W$_0$ values for benzene to pentacene and evGW values for hexacene.  The resulting comparison, across the entire acene series, is shown in Fig.~\ref{pes-ipes-gw}, with the rigid shift employed indicated on the figure. In perfect agreement with the findings of Ref.\ \onlinecite{sharifzadeh_quasiparticle_2012}, the rigid shift is very significant, with a combined PES and IPES shift of $\sim$1 eV. But after employing it, we find excellent agreement, in both energy position and line-shape, for all theoretical and experimental spectra across the entire acene series in a region up to $\sim$6~eV from the Fermi level.

\subsection{Neutral excitations in acene crystals} 
\label{sect:opt-properties}

Having discussed charged excitations, we now turn to analyzing lowest-energy singlet and triplet excitation energies in the acene crystals.
As in the previous sub-section, 
we use experimental lattice parameters in order to avoid errors associated with geometry.

\begin{figure}[h]
\includegraphics{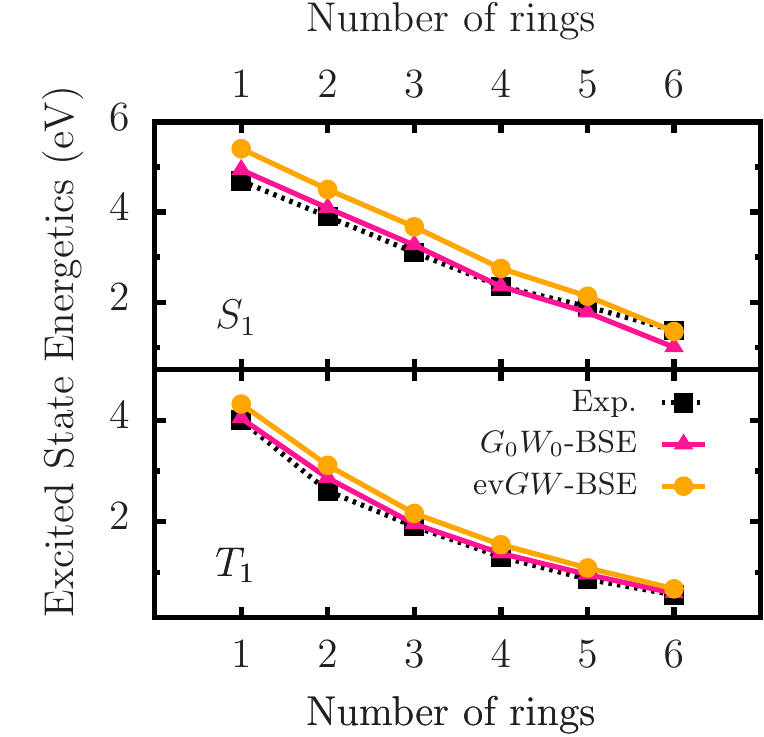}
\caption{(Color online)
Lowest lying excitation energies of acene the crystals, computed within the G$_0$W$_0$ (pink (medium-grey) lines (triangles)) and evGW-BSE orange ((grey) lines (circles)) approximations, compared to experimental data (black dotted lines (squares)), taken from Refs.~\onlinecite{hummer_oligoacene_2005,swiderek_electron_1990,*schnepp_electronic_1963,*grumstrup_enhanced_2010,*burgos_heterofission_1977,*hexacene_excitons,doering_electronic_1977} and references therein.
Experimental lattice constants have been used throughout.
Lowest singlet ($S_1$) and triplet ($T_1$) energies 
are shown at the top and bottom panels, respectively.}
\label{fig:g0w0-scgw}
\end{figure}

\begin{table}
\input{Figures/table1.tex}
\caption{
Lowest singlet, $S_1$, and triplet, $T_1$, excitation energies of the acene crystals, computed within the G$_0$W$_0$ and evGW-BSE approximations, compared to experimental data, taken from Refs.~\onlinecite{hummer_oligoacene_2005,swiderek_electron_1990,*schnepp_electronic_1963,*grumstrup_enhanced_2010,*burgos_heterofission_1977,*hexacene_excitons,doering_electronic_1977} and references therein.
Experimental lattice constants have been used in the calculations throughout.
All quantities are in eV.}
\label{table:g0w0-scgw}
\end{table}

Lowest neutral excitation energies, computed with both G$_0$W$_0$-BSE and evGW-BSE, are compared with experimental data in Fig.\ \ref{fig:g0w0-scgw}. The same comparison is also summarized in Table~\ref{table:g0w0-scgw}. Importantly, no significant temperature dependence of low-lying excitation energies is 
observed experimentally,\cite{helzel_temperature_2011,wilson_temperature-independent_2013} allowing for comparison to experiments performed at higher temperatures. Both calculations correctly predict the experimental quantum-size-effect trend, i.e., the decrease of $S_1$ and $T_1$ excitation energies with increasing acene size. However, for the $S_1$ excitations the computed slope is somewhat too large. Thus, the G$_0$W$_0$-BSE calculation overestimates experiment by 0.25 eV for benzene, agrees perfectly with experiment for tetracene, and underestimates experiment by 
0.4 eV for hexacene. Once again, we view this primarily as a starting point issue. For the larger acenes, the PBE gap is very small (only 0.2 eV for hexacene). Likely this results in increasingly worse over-screening, as in a simple model the dielectric constant is inversely proportional to the square of the quasi-particle gap.\cite{penn_wave-number-dependent_1962} This assertion is supported by the fact that for hexacene, a GW$_0$ approach,\cite{tiago_optical_2006,shishkin_self-consistent_2007} in which self-consistency in $G$ alone is performed, results in a singlet energy of $S_1=$1.06~eV, which is almost equivalent to the G$_0$W$_0$-BSE value of 1.00~eV.   
A starting point with a larger gap, as in evGW-BSE, leads to reduced screening and may therefore yield better neutral excited states for this system. In particular, the evGW-BSE value for hexacene is in perfect agreement with experiment. However, as with the charged excitations, evGW-BSE is not a panacea.\cite{shishkin_self-consistent_2007,shishkin_accurate_2007} it shifts the G$_0$W$_0$-BSE results by an almost uniform 0.3-0.4 eV, leading to an overestimate of $S_1$ for the smaller acenes.

The lowest triplet excitation energies, $T_1$, obtained from G$_0$W$_0$-BSE show a generally similar trend, but agree well with experiment for hexacene and show a modest overestimate for the smaller acenes, up to 0.3 eV for benzene and naphthalene. As with the singlet excitations, evGW-BSE calculations predict $T_1$ values in good agreement with experiments (within 0.2~eV) for pentacene and hexacene, but overestimate $T_1$ for the smaller systems, by as much as 0.5 eV for benzene. 

As mentioned above, all BSE calculations we have presented use the Tamm Dancoff approximation (TDA). 
The TDA was found to be accurate in describing the lowest-lying excitations of molecules and small silicon clusters,\cite{hirose_all-electron_2015,rocca_ab_2010,rocca_ab_2014} although this does not necessarily hold for larger chemical entities.\cite{gruning_exciton-plasmon_2009,ma_excited_2009}
The applicability of the TDA in three-dimensional solids has not been explored as much. 
Nevertheless, as an example we find for the tetracene crystal that the $S_1$ value obtained within G$_0$W$_0$-BSE is negligibly affected (by only 0.02 eV) upon relaxing the TDA.

To summarize, within our other assumptions -- a PBE starting point, the GPP model, and the Tamm-Dancoff approximation, G$_0$W$_0$-BSE is the optimal choice for the smaller acenes, up to tetracene, but evGW is better for the largest acenes -- hexacene and perhaps pentacene. 

\subsection{Effects of structure on charged and neutral excitations}
\label{sect:image-charge}

\begin{table}
\input{Figures/table5.tex}
\caption{Effect of structure on the k-point averaged fundamental gap, $\Delta_g$, along with the valence band width (vbw) and the conduction band width (cbw), all calculated in the G$_0$W$_0$ approximation. All values were obtained from lattice parameters fully relaxed within the LDA, PBE, DF1, DF2 and DF-cx functionals, as well as from  experimental parameters.\cite{bacon_crystallographic_1964,capelli_molecular_2006,
chaplot_structure_1982,schiefer_determination_2007,holmes_nature_1999} All energies are in eV.
}
\label{table:gw-structure}
\end{table}

Having discussed the need for adequate treatment of vdW interactions for predicting geometry and independently the accuracy of approximations within the GW and BSE schemes, we now turn to question of the sensitivity of the calculated excitations to structural parameters. 

We start by considering charged excitations obtained within the GW approximations and assessing their dependence on the geometry obtained from LDA, GGA, the three van-der-Waals functionals (DF1, DF2, and DF-cx) used in Section \ref{sect:geom-lattice} above, and experiment. The calculated k-point averaged G$_0$W$_0$-calculated fundamental gap, $\Delta_g$, along with the valence band width (vbw) and the conduction band width (cbw), for each of the geometries, is given in Table~\ref{table:gw-structure}. Hexacene is excluded here and below so that we can restrict our attention to G$_0$W$_0$ and avoid additional differences arising from comparison between G$_0$W$_0$ and evGW. 


As shown in Section~\ref{sect:geom-lattice} above, and in more detail in Appendix~\ref{appendix:vdws} below, lattice parameters increase with functional in the following sequence: LDA/DF1/DF2/DF-cx/PBE.
Interestingly, Table~\ref{table:gw-structure} shows that $\Delta_g$ follows the same trend, while vbw and cbw follow the opposite trend.  These trends can be rationalized as follows: the larger the lattice parameters, the smaller the inter-molecular hybridization and the smaller the band-width. Naturally, the smaller the hybridization, the larger the bandgap. However, quantitatively the change in band widths explains only part of the gap increase with increasing lattice parameters. A second effect is that the solid-state gap is renormalized from the much larger molecular gap (compare with the molecular gaps given in Table \ref{table:gas_phase} of Section~\ref{sect:gas-phase} because the neighboring molecules serve as a dielectric medium whose response creates a polarization field that reduces the gap.\cite{Sato81} As discussed in detail in Refs.\ \onlinecite{neaton_renormalization_2006,sharifzadeh_quasiparticle_2012,
Refaely13,*Refaely15}, this phenomenon, which is well-captured by GW calculation, itself depends on the unit-cell volume. This is because a larger inter-molecular separation reduces the polarization field and therefore the renormalization, thereby increasing the gap.

Finally, we note that the although GW gaps calculated from the experimental geometry are within 0.1 eV (0.2 eV for benzene) of those obtained with DF2 lattice parameters, it should be taken into account that this is due to the accidental agreement of zero-temperature DF2 volumes with room temperature experimental values (see Section~\ref{sect:geom-lattice}).


\begin{figure}[h!]
\includegraphics{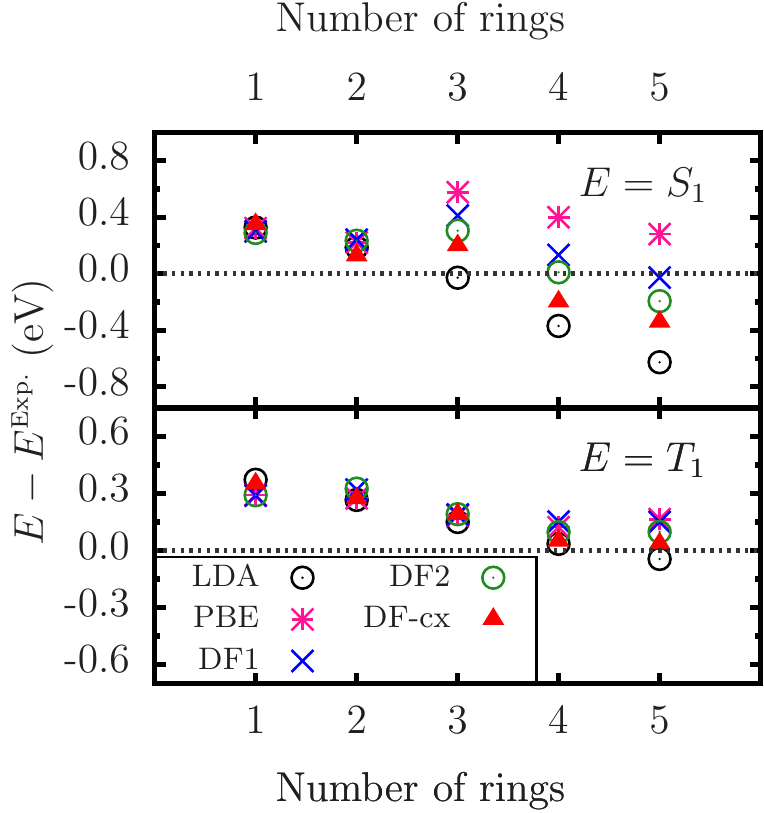}
\caption{(Color online)
Effect of structure on the lowest singlet~($S_1$) and triplet~($T_1$) excitation energies, calculated in the G$_0$W$_0$-BSE approximation, given as deviation from experimental data (see Refs. in Table.~\ref{table:g0w0-scgw}). All computed values were obtained from lattice parameters fully relaxed within the LDA (black empty circles), PBE (pink stars),
DF1 (blue crosses), DF2 (green empty squares), and
DF-cx (red triangles) functionals, and are given as differences from experimental values. All energies are in eV.
}
\label{fig:excited-states}
\end{figure}

\begin{table}
\input{Figures/table2.tex}
\caption{
Effect of structure on the lowest singlet~($S_1$) and triplet~($T_1$) excitation energies, calculated in the G$_0$W$_0$-BSE approximation. 
All computed values were obtained from lattice parameters fully relaxed within the LDA, PBE, DF1, DF2, and DF-cx functionals, as well as from experimental values. All energies are in eV.
For comparison, experimental values, taken from Refs.~\onlinecite{hummer_oligoacene_2005,swiderek_electron_1990,*schnepp_electronic_1963,*grumstrup_enhanced_2010,*burgos_heterofission_1977,*hexacene_excitons,doering_electronic_1977} and references therein, are also given.
Also given are the mean absolute error (MAE), defined as 
$\sum_i^{N_m} | X_i - X^{\scriptsize \mbox{Exp.}}_i |/  N_m$, with 
$X$ being the excitation energy, the maximum absolute error (MAXE), and the mean absolute percentge error (MA\%E), defined as 
MA\%E=$\sum_i^{N_m} | X_i - 
X^{\scriptsize \mbox{Exp.}}_i | / X^{\scriptsize \mbox{Exp.}}_i 
/  N_m 
\times 100 $.}
\label{table:excited-states}
\end{table}

We now turn to the discussion of structure on neutral excitations.
G$_0$W$_0$-BSE calculated low-lying excitations, based on the same geometries used in Table \ref{table:gw-structure} above, are given in Fig.~\ref{fig:excited-states} and in Table~\ref{table:excited-states}.
For comparison, Table~\ref{table:excited-states} also reports neutral excitations calculated using experimental lattice parameters and shown above to be in good agreement with experimental excitation energies (see section~\ref{sect:opt-properties}). As discussed above, no significant temperature dependence of low-lying excitation energies is 
observed,\cite{helzel_temperature_2011,wilson_temperature-independent_2013} which facilitates the comparison to experimental excitation energies measured at higher temperatures.

Clearly, the dependence of $T_1$ excitation energies on geometry is quite minimal ($\pm$0.1 eV at most across the entire acene series). The same is true for the $S_1$ excitations in the smaller acenes (benzene and naphthalene), but the dependence on geometry increases with acene size. For pentacene it is already quite significant, with the $S_1$ excitation values changing by 0.9 eV by switching from LDA to PBE geometry. As before, agreement with experiment is much improved by using DF-based geometry, with best results obtained using DF2 and DF-cx (with differences between the two being too small to be physically meaningful), based on which $S_1$ energies are found to be within 4\%-5\% of experimental values.
The remaining discrepancy may be due to terms not included in this work, such as zero-point and finite-temperature effects associated with lattice vibrations, as well as the remaining limitations of the GW-BSE approach in general and its approximations used here in particular.

The sensitivity of excitation energies to geometry, or lack thereof, is directly related to the degree of spatial localization of these states. 
For large acenes, notably pentacene, singlet states have been shown to extend over several molecules.\cite{tiago_ab_2003,hummer_oligoacene_2005,sharifzadeh_low-energy_2013,cudazzo_excitons_2012,*cudazzo_prb_2013} The degree of delocalization is larger for smaller unit cell volumes, an effect related to the increased inter-molecular hybridization \cite{sharifzadeh_relating_2015} Delocalization decreases the excitation energy, which is therefore larger the smaller the unit cell is. For short acenes this is a much smaller effect and indeed no significant structure-dependence is observed. Triplets, however, are always predominantly localized on a single monomer.\cite{sharifzadeh_low-energy_2013,sharifzadeh_relating_2015}, explaining their weak dependence on the geometry.  Owing to this negligible delocalization, triplet energies calculated in the gas- and solid-state are within 0.2~eV (compare with Table~\ref{table:energetics-gas-phase}),
i.e., the triplet is largely independent of the solid state environment.

Finally, we note that the calculated neutral excitation energies are not strongly affected by temperature, at least as reflected by the crystal structure used in our calculations.
For benzene, naphthalene, and anthracene, the calculated singlet and triplet energies do not depend significantly on geometry. 
For the larger acenes, singlet energies change by at most 0.15 eV, while triplet energies change by even less, when varying the volume by 
$\sim$3\% (similar to thermal expansion at room temperature). 
This agrees with the experimentally observed absence of significant temperature dependence of low-lying excitation energies.\cite{helzel_temperature_2011,wilson_temperature-independent_2013}

\section{Conclusions}
\label{conclusions}

In summary, we have studied the structure and excited state properties of the series of acene-based crystals, from benzene to hexacene, from first principles using vdW-corrected-DFT and MBPT. 
Both vdW-DF and pair-wise correction methods were found to predict lattice parameters in excellent agreement with experimental data.
We find that DF1 overestimates volumes but DF2 improves over DF1, consistent with the general trends for these functionals.
DF-cx further improves lattice parameters, with a residual discrepancy of $<$ 1\%). Furthermore, the relatively simple TS pair-wise approach performs as well as the best DF methods.

For acenes in the solid-state, charged excitations are generally well-described by the G$_0$W$_0$ method, but partial self-consistency -- in the form of the evGW method -- is needed for hexacene, likely owing to the PBE starting point employed in this study. The results are found to be sensitive to the geometry used owing to a combination of inter-molecular hybridization and polarization-induced level renormalization.
Neutral low-lying singlet and triplet excitation energies are generally well-described using the G$_0$W$_0$-BSE method. They are generally less sensitive to structure, except for the important case of singlet excitations in larger acenes. There, large structural sensitivity is found owing to significant delocalization of the singlet state.



Our study reveals the importance of an accurate account of dispersive interactions as a prerequisite to predictive calculations of excited states properties in the acene crystals. Furthermore, it suggests routes for predictive calculations, in which both structures and excited states are calculated entirely from first-principles, for broader classes of molecular solids.

\begin{acknowledgments}
T. Rangel thanks Marc Torrent and Muriel Delaveau for addressing technical issues in \textsc{ABINIT}, related to the calculation of a large number of bands needed for GW calculations.
This research was supported by the SciDAC Program on Excited State Phenomena in Energy Materials funded by the U.S. Department of Energy, Office of Basic Energy Sciences and of Advanced Scientific Computing Research, under Contract No. DE-AC02-05CH11231 at Lawrence Berkeley National Laboratory. 
Work at the Molecular Foundry was supported by the Office of Science, Office of Basic Energy Sciences, of the U.S. Department of Energy under Contract No. DE-AC02-05CH11231. 
This research used resources of the National Energy Research Scientific Computing Center, which is supported by the Office of Science of the U.S. Department of Energy.
Work in Sweden supported by the Swedish Research Council and the Chalmers Nanoscience Area of Advance.
Work in Israel was supported by the US-Israel Binational Science Foundation, the molecular foundry, and the computational resources of 
the National Energy Research Scientific Computing center.
\end{acknowledgments}

\appendix
\begin{figure*}[!h]
\includegraphics{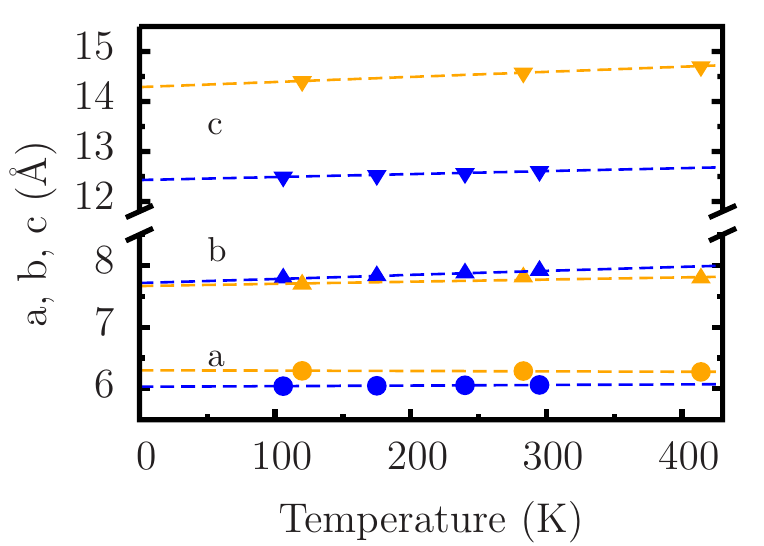}
\includegraphics{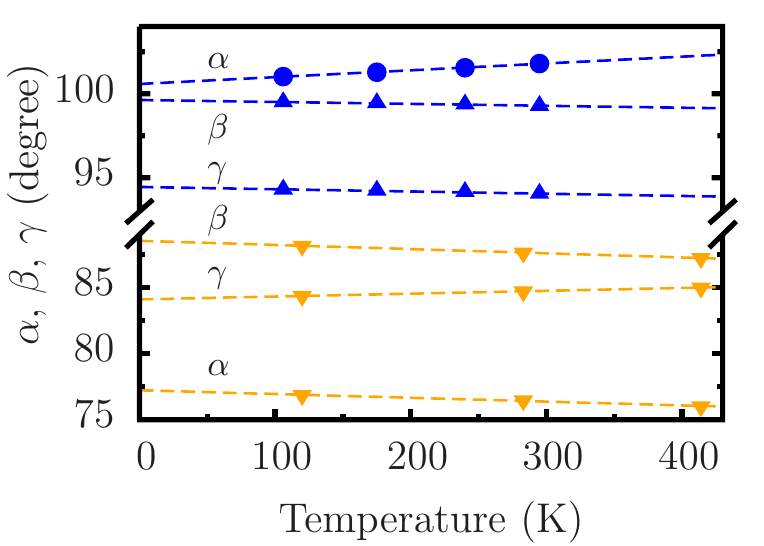}
\caption{(Color online) 
Extrapolation of unit cell geometry to 0 K:
Experimental lattice parameters and angles of 
tetracene~P$_1$~(blue) are extracted from Ref.~\onlinecite{haas_crystal_structure_2006} and those of
pentacene~P$_2$~(orange) are
extracted from Refs.~\onlinecite{siegrist_polymorph_2007},
also labeled as \textsc{PENCEN06} - 
\textsc{PENCEN08} in the CSD.
These are fitted to linear functions of the temperature (dashed lines).
The fits possess an average root mean square of
of $7\times10^{-3}$ and $0.05$~{\AA}
and $0.02$ and $0.05$~$^\circ$
for tetracene~P$_1$ and pentacene~P$_2$, respectively.
}
\label{fig:extrapolate}
\end{figure*}
\begin{figure*}
\begin{minipage}{0.48\linewidth}
\includegraphics[width=1.0\linewidth]{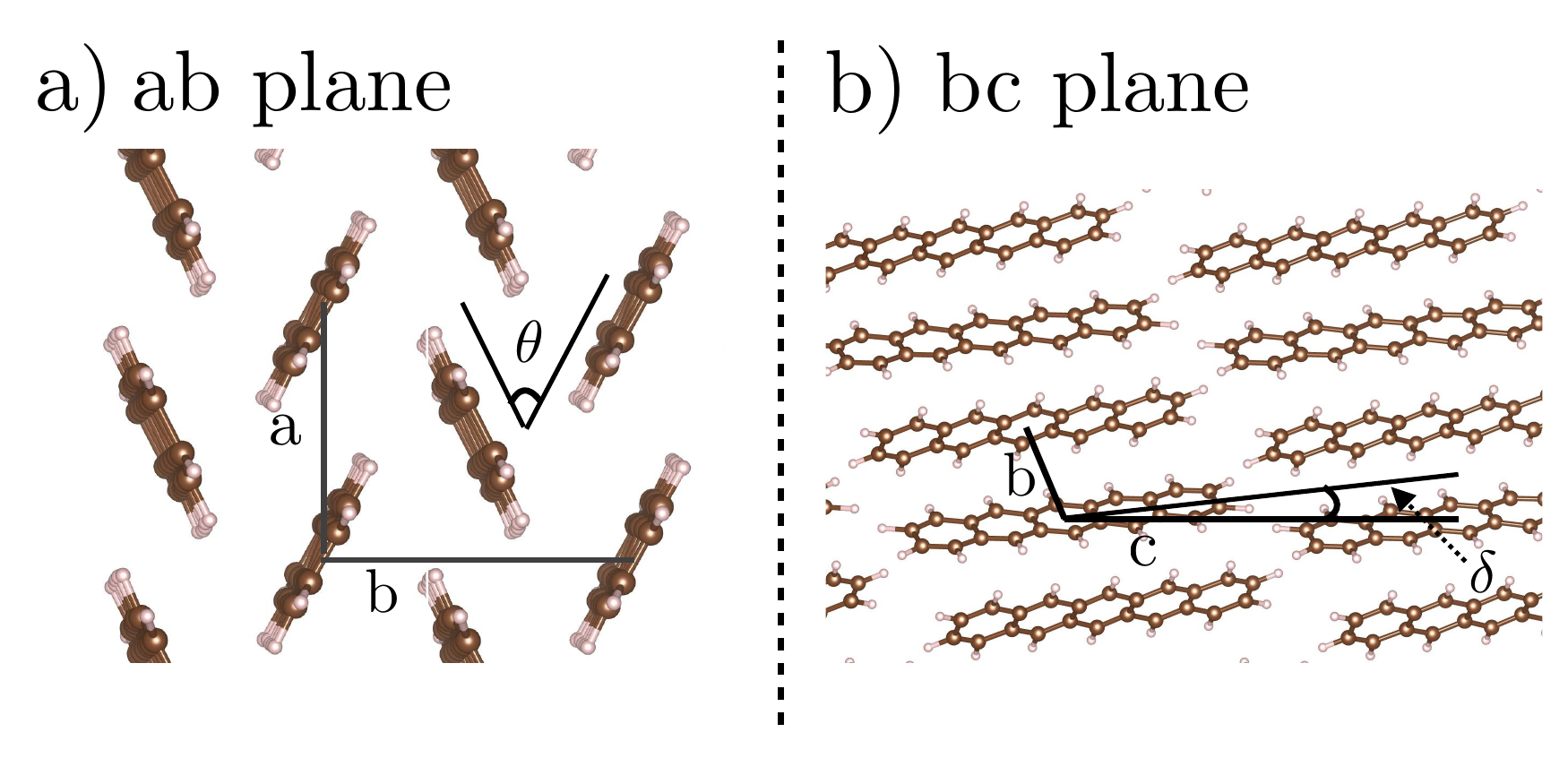}
\end{minipage}
\begin{minipage}{0.48\linewidth}
\begin{tabular}{@{}lccccc@{}}
\multicolumn{5}{l}{Pentacene:}\\
&& DF1 & DF2 & DF-cx & Exp. \\
\hline\\
\multirow{2}{*}{$P_1$} &
$\theta$ &
  50 &47 & 47 &53 \\
&$\delta$ &
  12 & 12 & 12 & 12
\\\hline\\
\multirow{2}{*}{$P_2$} &
$\theta$ &
 53   & 51   & 50  & 51\\
& $\delta$ &
 15   & 14   & 16   & 14
\\\hline\\
\multirow{2}{*}{$P_3$} &
$\theta$ & 
 55   & 55   & 53   & 54\\
& $\delta$ &
  5   & 4   & 4   & 4
\\\hline\\
\end{tabular}
\end{minipage}
\caption{Angles characterizing the 
pentacene herringbone structure:
(a) Along the ab plane, the herringbone angle, $\theta$,
is the angle between the two distinct molecules.
(b) Along the bc plane, $\delta$ is the angle between
the c axis and the long axis of the molecule.
(Right Table) Angles calculated by different DF methods are compared to experimental data from Refs.~\onlinecite{siegrist_polymorph_2007,campbell_crystal_1962,
schiefer_determination_2007},
labeled as \textsc{PENCEN}, \textsc{PENCEN06}, and \textsc{PENCEN10}
in the CSD, and measured at 205~K, 120~K and 293~K for 
pentacene P$_1$, P$_2$, and P$_3$, respectively. 
While $\theta$ is almost constant for all polymorphs ($\theta\sim50-54^\circ$), $\delta$  is not. We use Niggli unit-cell conventions and all angles are in degrees.
}
\label{fig:angles-pentacene}
\end{figure*}

\input{Figures/table3.tex}

\section{Structural properties: benchmarking of vdWs functionals}
\label{appendix:vdws}

In this Appendix we provide detailed information on the structural data obtained with different methods for the acene family of crystals.
As in the main text, we consider standard DFT methods (LDA and PBE) and different vdW methods: D2, TS, DF1, DF2 and DF-cx.

Throughout, we make use of CSD~\cite{csd} data to benchmark our results.
For the smallest acenes,  low temperature data ($T\leq14$~K) is available in the CSD under the entries \textsc{BENZEN14}, \textsc{NAPHTA31}, and
\textsc{ANTCEN16}, from Refs.~\onlinecite{david_crystal_1992,chaplot_structure_1982,
capelli_molecular_2006}.
Ref.~\onlinecite{craven_structure_1993} also reports low temperature data for benzene, consistent with the data of Ref.~\onlinecite{david_crystal_1992}.
For tetracene~P1 and pentacene-P2, we extrapolate experimental data from Refs.~\onlinecite{haas_crystal_structure_2006,siegrist_polymorph_2007} to zero Kelvin, as shown in Fig.~\ref{fig:extrapolate}.
Note that we assign the tetracene structures of Ref.~\onlinecite{haas_crystal_structure_2006} to its P1 polymorph.\footnote{The tetracene lattice parameters reported in the Ph.D. thesis of Haas\cite{haas_crystal_structure_2006} at $175$~K almost perfectly match P$_1$ at $180$~K as reported in Ref.~\onlinecite{venuti_phonons_2004} (the volumes agree within 0.5\%). 
Further, by cooling the sample from $175$~K to $106$~K, Haas reports a slight reduction in volume, as shown in Figure~\ref{fig:extrapolate} of our manuscript. 
Hence, we can assign Haas's structures to the P$_1$ polymorph. 
Note that the P$_2$ polymorph has an appreciably smaller c lattice-vector (12.3~\AA) than the P$_1$ phase ($\sim$12.5~\AA).}
For other pentacene polymorphs and hexacene, in the absence of sufficient low-temperature data that would allow for extrapolation to $0$~K, we compare to the lowest-temperature experimental data available from Refs.~\onlinecite{campbell_crystal_1962,watanabe_synthesis_2012,
schiefer_determination_2007}, also found in the CSD as \textsc{PENCEN}, \textsc{PENCEN10}, and \textsc{ZZZDKE01}.
We emphasize that only by extrapolating experimental data to $0$~K do we observe consistent trends in the comparison of our relaxed geometries for the various DFT methods used here. 
In the main text, we have also compared our data to experimental cohesive energies. 
These are taken from Ref.~\onlinecite{reilly_understanding_2013}, in which temperature contributions have been removed.
A complete set of experimental and calculated  lattice parameters and cohesive energies is given in Tables~\ref{table:volumes} and \ref{table:cohesive}, respectively. Lattice parameters are usually found in literature following old conventions.
However, recent data use the so called Niggli~\cite{santoro_determination_1970} (or reduced-) lattice parameters.
For completeness, we present both conventions in Table~\ref{table:volumes}.
Finally, in Fig.~\ref{fig:angles-pentacene} we present a comparison of theory and experiment for the angles that characterize the herringbone structure in the three pentacene polymorphs. 
Here, all DF approximations predict angles in good agreement with experiment. At the experimental resolution and temperature, we cannot conclude definitively which DF version performs best for angle prediction, but see no reason for trends different from those reported in the main text.

\pagebreak

\LTcapwidth=\textwidth 
\begin{longtable*}{p{1.8cm}p{1.5cm}p{1.5cm}p{1.5cm}p{1.5cm}p{1.5cm}p{1.5cm}p{1.5cm}p{1.5cm}p{1.5cm}p{1.5cm}}
\caption{ Structural parameters for the acene crystal series. 
Lattice parameters (using old conventions) $a$, $b$ and $c$ (in \AA), lattice angles $\alpha$, $\beta$ and $\gamma$ (in degrees) and unit cell volumes $\Omega$ (in \AA$^3$)  are calculated from force and stress relaxation within different DFT approximations.
Niggli lattice parameters~($'$) are also shown wherever they differ from old-convention ones.
Experimental lattice parameters 
measured at low temperature $T$ are also shown.
For pentacene, three different polymorphs are considered, labeled as 
P$_1$, P$_2$ and P$_3$.
$*$ Where possible, lattice parameters are extrapolated to 0~K (see text for details).
Mean absolute errors~(MAE) and mean absolute percentage errors~(MA\%E) with respect to experimental lattice parameters are also shown:
In the error evaluation, 
we only considered experimental data with $T \leq 16$~K or data extrapolated to 0~K. For brevity, symmetry-imposed
angles are omitted.
TS data taken from Ref.~\onlinecite{otero-de-la-roza_benchmark_2012} for benzene and from Ref.~\onlinecite{david_crystal_1992,chaplot_structure_1982,
capelli_molecular_2006,schatschneider_understanding_2013}
for longer acenes. 
Experimental data are taken from Refs.~\onlinecite{david_crystal_1992,chaplot_structure_1982,
capelli_molecular_2006,haas_crystal_structure_2006,campbell_crystal_1962,
schiefer_determination_2007,watanabe_synthesis_2012}.
}
\label{table:volumes}
\endfirsthead
\hline\hline\\
\input{Figures/table4.tex}
\hline
\end{longtable*}

\bibliography{paper}

\end{document}

%% file: Figures/table6.tex
\begin{ruledtabular}
\begin{tabular}{c|ccc}
&&&\\
\# of &
\multicolumn{3}{c}{$\Delta_g$} \\
rings &  
G$_0$W$_0$ & evGW & Exp.\\
1 &  7.3 & 8.2 & 7.6 -- 8.0 \\
2 &  5.5 & 6.1 & 5.0 -- 5.5 \\
3 &  4.0 & 4.5 & 3.9 -- 4.2 \\
4 &  2.9 & 3.5 & 2.9 -- 3.4 \\
5 &  2.2 & 2.8 & 2.2 -- 2.4 \\ 
6 &  1.3 & 1.8 & \\
\end{tabular}
\end{ruledtabular}

%% file: Figures/table1.tex
\begin{ruledtabular}
\begin{tabular}{c|ccc|ccc}
&&&&&&\\
\# of &
\multicolumn{3}{c|}{$T_1$} &
\multicolumn{3}{c}{$S_1$} \\
rings &  
G$_0$W$_0$ & evGW & Exp. &
G$_0$W$_0$ & evGW & Exp.\\
1 &  4.1 &  4.3 &  3.8  &  5.0 & 5.4 & 4.7 \\
2 &  2.9 &  3.1 &  2.6  &  4.2 & 4.5 & 3.9  \\
3 &  2.0 &  2.2 &  1.9  &  3.3 & 3.7 & 3.1  \\
4 &  1.4 &  1.5 &  1.3  &  2.4 & 2.8 & 2.4 \\
5 &  1.0 &  1.1 &  0.9 &  1.8 & 2.1 & 1.9  \\
6 &  0.6 &  0.7 &  0.6 &  1.0 & 1.4 & 1.4 \\
\end{tabular}
\end{ruledtabular}

%% file: Figures/table5.tex
\begin{ruledtabular}
\begin{tabular}{cccccccc}
\\
& \multicolumn{6}{c}{G$_0$W$_0$}\\
& \multicolumn{6}{c}{ using lattice-parameters from:}\\
 &  LDA & PBE & DF1 & DF2 & DF-cx & Exp. \\
\\\multicolumn{6}{l}{Benzene}  &  \\ 
 $\Delta_g$ &  6.6 &  7.9 &  7.3 & 7.1  & 7.1 &   7.3 \\
 vbw        &  0.7 &  0.2 &  0.4 & 0.5  & 0.5 &   0.4 \\
 cbw        &  0.7 &  0.4 &  0.5 & 0.5  & 0.5 &   0.5 \\
\\\multicolumn{6}{l}{Naphthalene}& \\ 
 $\Delta_g$ &  4.9  & 6.0  &  5.6  & 5.5  & 5.3  &  5.5 \\
 vbw        &  0.7  & 0.2  &  0.4  & 0.5  & 0.5  &  0.5\\
 cbw        &  0.7  & 0.2  &  0.3  & 0.3  & 0.4  &  0.3 \\
\\\multicolumn{6}{l}{Anthracene} &  \\  
 $\Delta_g$  & 3.6 & 4.5  & 4.3  & 4.1  & 4.0  & 4.0 \\
 vbw         & 0.6 & 0.2  & 0.3  & 0.4  & 0.4  & 0.3  \\
 cbw         & 1.0 & 0.3  & 0.5  & 0.6  & 0.7  & 0.6  \\
\\\multicolumn{6}{l}{Tetracene}   \\ 
 $\Delta_g$  & 2.4  & 3.6 & 3.1  & 2.9  & 2.7  &  2.9 \\
 vbw         & 0.7  & 0.1 & 0.3  & 0.4  & 0.6  &  0.4 \\
 cbw         & 0.9  & 0.3 & 0.5  & 0.7  & 0.8  &  0.7 \\
\\\multicolumn{6}{l}{Pentacene P3} \\ 
 $\Delta_g$ &  1.5  & 2.9  &2.3  & 2.1  & 1.8  & 2.2  \\
 vbw        &  1.2  & 0.3  &0.6  & 0.7  & 0.9  & 0.7  \\
 cbw        &  1.1  & 0.3  &0.6  & 0.7  & 0.9  & 0.7  \\
\end{tabular}
\end{ruledtabular}

%% file: Figures/table2.tex
\begin{ruledtabular}
\begin{tabular}{cc|cccccc|c}
&&&&&&&&\\
&& \multicolumn{6}{c|}{G$_0$W$_0$-BSE} & \multirow{3}{*}{Exp.}\\
&& \multicolumn{6}{c|}{ using lattice-parameters from:}&\\
 && LDA & PBE & DF1 & DF2 & DF-cx & Exp. \\
\hline
&&&&&&&\\\multicolumn{2}{l|}{Benzene}&&&&&&\\
& $S_1$ &  5.1 &  5.0 &  5.0 &  5.0 &  5.0 &  5.0 &  4.7 \\
& $T_1$ &  4.2 &  4.0 &  4.1 &  4.0 &  4.1 &  4.1 &  3.7 \\
\hline
&&&&&&&\\\multicolumn{2}{l|}{Naphthalene}&&&&&&\\
& $S_1$ &  4.1 &  4.2 &  4.2 &  4.2 &  4.1 &  4.2 &  3.9 \\
& $T_1$ &  2.9 &  2.9 &  2.9 &  2.9 &  2.9 &  2.9 &  2.6 \\
\hline
&&&&&&&\\\multicolumn{2}{l|}{Anthracene}&&&&&&\\
& $S_1$ &  3.1 &  3.6 &  3.5 &  3.4 &  3.3 &  3.4 &  3.1 \\
& $T_1$ &  2.0 &  2.0 &  2.1 &  2.0 &  2.0 &  2.0 &  1.9 \\
\hline
&&&&&&&\\\multicolumn{2}{l|}{Tetracene}&&&&&&\\
& $S_1$ &  2.0 &  2.9 &  2.5 &  2.4 &  2.2 &  2.4 &  2.4 \\
& $T_1$ &  1.3 &  1.4 &  1.4 &  1.4 &  1.3 &  1.4 &  1.3 \\
\hline
&&&&&&&\\\multicolumn{2}{l|}{Pentacene P3}&&&&&&\\
& $S_1$ &  1.3 &  2.1 &  1.9 &  1.7 &  1.5 &  1.7 &  1.9 \\
& $T_1$ &  0.8 &  0.9 &  0.9 &  0.9 &  0.8 &  0.9 &  0.9 \\
&&&&&&&&\\ \multirow{3}{*}{$S_1$} & MAE&  0.3&  0.4&  0.2&  0.2&  0.3&  0.2\\
 & MA\%E & 12& 12&  7&  7& 10&  7\\
 & MAXE&  0.6&  0.5&  0.4&  0.3&  0.4&  0.4
\\&&&&&&&&\\ \multirow{3}{*}{$T_1$} & MAE&  0.2&  0.2&  0.2&  0.2&  0.2&  0.2\\
 & MA\%E &  8&  9& 10&  7&  6&  8\\
 & MAXE&  0.5&  0.4&  0.4&  0.3&  0.4&  0.4\\
\end{tabular}
\end{ruledtabular}

%% file: Figures/table3.tex
\begin{table}
\setlength\extrarowheight{1pt}
\begin{ruledtabular}
\begin{tabular}{ccccccccc}
&\multicolumn{8}{c}{Cohesive energy~[eV]}\\
 &  LDA & PBE & D2 & TS$^*$ & DF1 & DF2 & DF-cx & Exp.  \\
Benzene&  0.59 & 0.12 & 0.73 & 0.69 & 0.64 & 0.60 & 0.61 & 0.52 \\
Naphthalene&  0.76 & 0.15 & 1.16 & 1.04 & 0.93 & 0.86 & 0.92 & 0.82 \\
Anthracene&  0.97 & 0.19 & 1.61 & 1.39 & 1.24 & 1.16 & 1.23 & 1.13 \\
Tetracene&  1.21 & 0.25 & 2.10 & & 1.56 & 1.42 & 1.56 & \\
Pentacene P$_1$&  1.46 & 0.30 & 2.61 & & 1.88 & 1.76 & 1.87 & \\
Pentacene P$_2$&  1.48 & 0.30 & 2.63 & & 1.88 & 1.76 & 1.92 & \\
Pentacene P$_3$&  1.42 & 0.31 & 2.61 & & 1.88 & 1.79 & 1.87 & \\
Hexacene&  1.82 & 0.36 & 2.18 & & 2.21 & 2.09 & 2.30 & \\
\\ MAE [\AA]&  0.09& 0.66& 0.35& 0.22& 0.11& 0.05& 0.10\\
MA\%E &  11& 80& 42& 28& 16&  8& 13\\
\end{tabular}
\end{ruledtabular}
\caption{ 
Cohesive energies of the acenes.
Calculated 
($E_{\scriptsize \mbox{coh.}}$)
and experimental 
($E_{\scriptsize \mbox{coh.}}^{\scriptsize \mbox{Exp.}}$)
cohesive energies are tabulated.
Experimental cohesive energies are taken from Ref.~\onlinecite{reilly_understanding_2013}.
MAE and MA\%E are shown for all functionals:
MAE=$\sum_i^{N_m} 
|E^{\scriptsize \mbox{Exp.}}_{\scriptsize \mbox{coh.},i}
- E_{\scriptsize \mbox{coh.},i}| / N_m $
and 
MA\%E=$\sum_i^{N_m} |E^{\scriptsize \mbox{Exp.}}_{\scriptsize \mbox{coh.},i}
- E_{\scriptsize \mbox{coh.},i}
|/ E_{\scriptsize \mbox{coh.},i} / N_m \times 100$,
where $N_m$ is the total number of crystals.
$^*$ TS data taken from Ref.~\onlinecite{reilly_understanding_2013}.
}
\label{table:cohesive}
\end{table}

%% file: Figures/table4.tex
&  LDA & PBE &D2 & TS$^*$ & DF1 & DF2 & DF-cx & \multicolumn{2}{c}{Exp. }  \\
\\\multicolumn{8}{l}{Benzene}  & $T=4$~K\\
$ a $ & 6.37 & 7.46 & 6.43 & 6.95 & 7.01 & 6.87 & 6.73 & 6.70 \\
$ b $ & 7.06 & 8.11 & 7.12 & 7.58 & 7.49 & 7.39 & 7.40 & 7.36 \\
$ c $ & 8.96 & 9.93 & 9.05 & 9.51 & 9.89 & 9.41 & 9.49 & 9.37 \\
$ \Omega $ & 403.0 & 601.1 & 414.4 & 500.8 & 518.4 & 477.9 & 473.0 & 461.8 \\
\\\multicolumn{8}{l}{Naphthalene}  & $T=5$~K\\
$ a $ & 7.74 & 8.99 & 7.79 & 8.12 & 8.48 & 8.22 & 8.06 & 8.08 \\
$ b $ & 5.76 & 6.31 & 5.79 & 5.90 & 6.06 & 5.97 & 5.91 & 5.93 \\
$ c $ & 8.37 & 9.13 & 8.44 & 8.65 & 8.79 & 8.57 & 8.75 & 8.63 \\
$ \beta $ & 125.5 & 122.1 & 125.3 & 124.2 & 123.3 & 122.9 & 124.4 & 124.7 \\
$ \Omega $ & 304.0 & 438.9 & 310.4 & 342.2 & 377.9 & 353.3 & 344.4 & 340.4 \\
$ a' $ & 5.76 & 6.31 & 5.79 & 5.90 & 6.06 & 5.97 & 5.91 & 5.93 \\
$ b' $ & 7.40 & 8.77 & 7.47 & 7.85 & 8.21 & 8.03 & 7.87 & 7.78 \\
$ c' $ & 7.74 & 8.99 & 8.44 & 8.12 & 8.48 & 8.22 & 8.06 & 8.08 \\
$ \alpha' $ & 112.9 & 118.1 & 121.7 & 114.5 & 116.4 & 116.4 & 113.4 & 114.1 \\
$ \Omega' $ & 304.0 & 438.9 & 310.4 & 342.2 & 377.9 & 353.3 & 344.4 & 340.4 \\
\\\multicolumn{8}{l}{Anthracene}  & $T=16$~K\\
$ a $ & 8.10 & 9.63 & 8.13 & 8.40 & 8.75 & 8.56 & 8.38 & 8.37 \\
$ b $ & 5.80 & 6.35 & 5.85 & 5.91 & 6.12 & 6.02 & 5.96 & 6.00 \\
$ c $ & 10.82 & 10.14 & 10.89 & 11.12 & 11.11 & 11.07 & 11.23 & 11.12 \\
$ \beta $ & 126.5 & 109.7 & 126.5 & 125.2 & 123.4 & 124.2 & 125.6 & 125.4 \\
$ \Omega $ & 408.9 & 584.4 & 416.6 & 451.0 & 496.7 & 471.2 & 456.5 & 455.2 \\
$ a' $ & 5.80 & 6.35 & 5.85 & 5.91 & 6.12 & 6.02 & 5.96 & 6.00 \\
$ b' $ & 8.10 & 9.63 & 8.13 & 8.40 & 8.75 & 8.56 & 8.38 & 8.37 \\
$ c' $ & 8.86 & 10.14 & 8.91 & 9.31 & 9.65 & 9.44 & 9.32 & 9.26 \\
$ \alpha' $ & 100.9 & 109.7 & 100.7 & 102.4 & 105.9 & 104.3 & 101.4 & 102.0 \\
$ \Omega' $ & 408.9 & 584.4 & 416.6 & 451.0 & 496.7 & 471.2 & 456.5 & 455.2 \\
\\\multicolumn{8}{l}{Tetracene~P$_1$}  & $T=0$~K$^*$ & $T=106$~K\\
$ a $ & 5.89 & 6.35 & 5.93 & 6.05 & 6.15 & 6.05 & 6.05 & 6.03 & 6.04 \\
$ b $ & 7.43 & 9.26 & 7.43 & 7.71 & 8.21 & 7.92 & 7.69 & 7.71 & 7.79 \\
$ c $ & 12.45 & 13.60 & 12.55 & 13.03 & 13.34 & 13.16 & 12.93 & 12.88 & 12.95 \\
$ \alpha $ &  78.4 &  72.6 &  78.7 &  77.7 &  75.1 &  75.6 &  78.0 &  77.6 &  77.3 \\
$ \beta $ &  72.8 &  71.5 &  72.5 &  71.9 &  71.2 &  72.0 &  72.6 &  72.1 &  72.1 \\
$ \gamma $ &  85.1 &  86.2 &  85.1 &  85.7 &  86.4 &  86.0 &  85.5 &  85.5 &  85.7 \\
$ \Omega $ & 510.0 & 723.2 & 516.6 & 564.1 & 616.1 & 579.9 & 561.3 & 557.7 & 566.1 \\
$ a' $ & 5.89 & 6.35 & 5.93 & 6.05 & 6.15 & 6.05 & 6.05 & 6.03 & 6.04 \\
$ b' $ & 7.43 & 9.26 & 7.43 & 7.71 & 8.21 & 7.92 & 7.69 & 7.71 & 7.68 \\
$ c' $ & 12.09 & 13.06 & 12.16 & 12.53 & 12.76 & 12.67 & 12.53 & 12.43 & 12.50 \\
$ \alpha' $ &  99.5 & 106.2 &  99.2 & 100.7 & 103.8 & 103.0 & 100.2 & 100.6 & 101.0 \\
$ \beta' $ & 100.5 &  99.0 & 100.2 &  99.2 &  98.3 &  99.0 & 100.1 &  99.6 &  99.5 \\
$ \gamma' $ &  94.9 &  93.8 &  94.9 &  94.3 &  93.6 &  94.0 &  94.5 &  94.5 &  94.3 \\
$ \Omega' $ & 510.0 & 723.2 & 516.6 & 564.1 & 616.1 & 579.9 & 561.3 & 557.7 & 566.1 \\
\\\multicolumn{8}{l}{Pentacene P$_1$}  & $T=295$~K\\
$ a $ & 7.37 & 9.14 & 7.36 & 7.66 & 8.17 & 7.90 & 7.59 & 7.90 \\
$ b $ & 5.97 & 6.33 & 6.00 & 6.04 & 6.16 & 6.08 & 6.07 & 6.06 \\
$ c $ & 15.54 & 16.13 & 15.64 & 15.85 & 15.99 & 15.83 & 15.98 & 16.01 \\
$ \alpha $ & 103.8 & 101.6 & 103.5 && 100.7 & 101.1 & 102.0 & 101.9 \\
$ \beta $ & 113.9 & 110.6 & 114.0 && 111.2 & 111.8 & 112.5 & 112.6 \\
$ \gamma $ &  84.7 &  86.2 &  84.7 &&  85.8 &  86.1 &  85.7 &  85.8 \\
$ \Omega $ & 607.0 & 856.1 & 613.5 & 664.1 & 736.8 & 692.7 & 666.0 & 692.4 \\
$ a' $ & 5.97 & 6.33 & 6.00 && 6.16 & 6.08 & 6.07 & 6.06 \\
$ b' $ & 7.37 & 9.14 & 7.36 && 8.17 & 7.90 & 7.59 & 7.90 \\
$ c' $ & 14.23 & 15.49 & 14.33 && 15.09 & 14.84 & 14.84 & 14.88 \\
$ \alpha' $ &  83.5 & 102.9 &  94.0 &&  99.1 &  97.9 &  95.7 &  96.7 \\
$ \beta' $ &  78.1 &  99.8 & 102.0 &&  99.0 &  99.7 & 100.7 & 100.5 \\
$ \gamma' $ &  84.7 &  93.8 &  95.3 &&  94.2 &  93.9 &  94.3 &  94.2 \\
$ \Omega' $ & 607.0 & 856.1 & 613.5 & 664.1 & 736.8 & 692.7 & 666.0 & 692.4 \\
\\\multicolumn{8}{l}{Pentacene P$_2$}  & $T=0$~K$^*$ & $T=120$~K \\
$ a $ & 6.18 & 6.52 & 6.25 & 6.13 & 6.45 & 6.33 & 6.29 & 6.30 & 6.29 \\
$ b $ & 7.27 & 8.91 & 7.24 & 7.68 & 8.07 & 7.81 & 7.52 & 7.67 & 7.69 \\
$ c $ & 13.80 & 15.16 & 13.85 & 14.53 & 14.69 & 14.49 & 14.35 & 14.29 & 14.41 \\
$ \alpha $ &  78.1 &  71.4 &  78.5 &  77.3 &  74.7 &  76.2 &  77.8 &  77.2 &  76.9 \\
$ \beta $ &  89.4 &  87.6 &  89.3 &  87.4 &  88.5 &  88.1 &  88.7 &  88.5 &  88.2 \\
$ \gamma $ &  83.7 &  84.9 &  83.5 &  84.7 &  84.8 &  84.6 &  84.1 &  84.1 &  84.4 \\
$ \Omega $ & 603.0 & 830.7 & 609.8 & 663.9 & 734.0 & 693.5 & 660.3 & 669.4 & 674.7 \\
\\\multicolumn{8}{l}{Pentacene P$_3$}  & $T=293$~K\\
$ a $ & 5.71 & 6.16 & 5.78 & 6.65 & 6.04 & 5.88 & 5.92 & 5.96 \\
$ b $ & 7.05 & 8.87 & 7.01 & 6.92 & 7.94 & 7.76 & 7.36 & 7.60 \\
$ c $ & 15.29 & 15.95 & 15.41 & 16.27 & 15.54 & 15.53 & 15.68 & 15.61 \\
$ \alpha $ &  82.5 &  81.4 &  82.8 &&  81.9 &  80.8 &  81.8 &  81.2 \\
$ \beta $ &  89.8 &  87.7 &  90.0 &&  87.4 &  87.9 &  87.2 &  86.6 \\
$ \gamma $ &  90.0 &  90.0 &  90.0 &&  89.4 &  89.7 &  89.7 &  89.8 \\
$ \Omega $ & 610.9 & 861.1 & 619.2 & 746.3 & 737.2 & 699.0 & 675.9 & 697.0 \\
\\\multicolumn{8}{l}{Hexacene}  & $T=123$~K\\
$ a $ & 6.47 & 6.60 & 6.48 && 6.43 & 6.34 & 6.61 & 6.31 \\
$ b $ & 6.85 & 9.10 & 6.85 && 8.04 & 7.84 & 7.05 & 7.70 \\
$ c $ & 15.64 & 17.30 & 15.71 && 16.79 & 16.49 & 16.14 & 16.48 \\
$ \alpha $ &  95.2 &  75.1 &  95.1 && 101.1 &  99.9 &  95.8 &  98.8 \\
$ \beta $ &  92.0 &  85.4 &  92.1 &&  90.6 &  91.3 &  91.5 &  91.2 \\
$ \gamma $ &  97.1 &  84.9 &  97.2 &&  95.4 &  95.5 &  96.8 &  95.8 \\
$ \Omega $ & 684.0 & 998.4 & 688.8 && 848.3 & 803.3 & 743.3 & 785.9 \\
\\ MAE [\AA]& 0.29& 0.76& 0.25& 0.09& 0.28& 0.11& 0.06\\
MA\%E&  3&  9&  3&  1&  3&  1&  1\\